\documentclass[manuscript,11pt]{acmart}

\usepackage{amsmath,amsthm,amsfonts,braket}
\usepackage{algorithmicx}
\usepackage{graphicx}
\usepackage{textcomp}
\usepackage{xcolor}
\usepackage{algorithm,balance}
\usepackage{pgfplots}
\usepackage{array}
\usepackage[noend]{algpseudocode}
\newtheorem{theorem}{Theorem}
\usepackage{subfigure}

\usepackage{epsfig}
\usepackage{algorithm}

\newcommand{\remove}[1]{}

\newcommand{\mysection}[1]{\vspace{-.03in}\section{#1}\vspace{-.00in}}

\newtheorem{lemma}{Lemma}
\newtheorem{corollary}{Corollary}

\newcommand{\ls}[1]
   {\dimen0=\fontdimen6\the\font
    \lineskip=#1\dimen0
    \advance\lineskip.5\fontdimen5\the\font
    \advance\lineskip-\dimen0
    \lineskiplimit=.9\lineskip
    \baselineskip=\lineskip
    \advance\baselineskip\dimen0
    \normallineskip\lineskip
    \normallineskiplimit\lineskiplimit
    \normalbaselineskip\baselineskip
    \ignorespaces
   }

\newcommand{\dubfigsingle}[6]{
\begin{figure*}[t]
\centerline{
    \begin{minipage}{0.42\textwidth}
      \begin{center}
        \leavevmode
        \setlength{\epsfxsize}{0.85\textwidth}
        \setlength{\epsfysize}{0.85\textwidth}
        \epsffile{\Figdir#1}
       \newline{\small (a) #2}
      \end{center}
    \end{minipage}
    \begin{minipage}{0.42\textwidth}
      \begin{center}
        \leavevmode
        \setlength{\epsfxsize}{0.85\textwidth}
        \setlength{\epsfysize}{0.85\textwidth}
        \epsffile{\Figdir#3}
       \newline{\small (b) #4}
      \end{center}
    \end{minipage}
} \caption{#5}\label{fig:#6}
\end{figure*}
}

\newcommand{\Figdir}{}

\newcommand{\kb}[1]{\ket{#1}\bra{#1}}

\newcommand\note[3]{{\textcolor{#1}{[\textsf{#2}: #3]}}}

\newcommand{\bing}[1]{\note{red}{BW}{#1}}
\newcommand{\vicky}[1]{\note{brown}{VM}{#1}}

\algnewcommand\algorithmicforeach{\textbf{for each}}
\algdef{S}[FOR]{ForEach}[1]{\algorithmicforeach\ #1\ \algorithmicdo}

\def\BibTeX{{\rm B\kern-.05em{\sc i\kern-.025em b}\kern-.08em
    T\kern-.1667em\lower.7ex\hbox{E}\kern-.125emX}}
\begin{document}


\title{Dynamic Routing for Quantum Key Distribution Networks}

\author{Omar Amer}
\email{omar.amer@jpmchase.com}
\affiliation{
\institution{Global Technology Applied Research, JPMorgan Chase Bank}
 \streetaddress{237 Park Ave}
 \city{New York City}
 \state{New York}
 \country{USA}
}
\affiliation{%
 \institution{Computer Science \& Engineering Department, University of Connecticut}
 \streetaddress{371 Fairfield Way}
 \city{Storrs}
 \state{CT}
 \country{USA}
}
\author{Walter O. Krawec}
\email{walter.krawec@uconn.edu}
\affiliation{%
\institution{Computer Science \& Engineering Department, University of Connecticut}
\streetaddress{371 Fairfield Way}
\city{Storrs}
\state{CT }
\country{USA}
}

\author{Victoria U. Manfredi}
\email{vumanfredi@wesleyan.edu}
\affiliation{%
\institution{Department of Mathematics and Computer Science, Wesleyan University}
\streetaddress{265 Church Street}
\city{Middletown}
\state{CT }
\country{USA}
}

\author{Bing Wang}
\email{bing@uconn.edu}
\affiliation{%
\institution{Computer Science \& Engineering Department, University of Connecticut}
\streetaddress{371 Fairfield Way}
\city{Storrs}
\state{CT }
\country{USA}
}

\pagenumbering{arabic}

\begin{abstract}
In this paper, we consider quantum key distribution (QKD) in a quantum network with both quantum repeaters and a small number of trusted nodes. In contrast to current QKD networks with only trusted nodes and the true Quantum Internet with only quantum repeaters, such networks represent a middle ground, serving as near-future QKD networks.  In this setting, QKD can be efficiently and practically deployed, while providing insights for the future true Quantum Internet. To significantly improve the key generation efficiency in such networks, we develop a new dynamic routing strategy that makes routing decisions based on the current network state, as well as evaluate various classical/quantum post-processing techniques. Using simulations, we show that our dynamic routing strategy can improve the key rate between two users significantly in settings with asymmetric trusted node placement. The post-processing techniques can also increase key rates in high noise scenarios. Furthermore, combining the dynamic routing strategy with the post-processing techniques can further improve the overall performance of the QKD network.
\end{abstract}

\maketitle
\pagenumbering{arabic}

\section{Introduction}

Quantum key distribution (QKD) allows two parties, typically referred to as Alice ($A$) and Bob ($B$), to establish a shared secret key, secure against computationally unbounded adversaries.  This is in contrast to classical key distribution protocols, where computational assumptions are always required to prove security.  Despite the great potential of QKD, however, numerous challenges remain.  Of particular importance is improving the overall key generation rate while also supporting larger distances between parties.  Typically, efficiency of a QKD protocol drops as the distance between parties and the noise in the channel increases.  Distance induced by loss in the channel is also a limiting factor as we know from the PLOB bound \cite{pirandola2017fundamental}, which gives a hard-limit to the capabilities of a QKD system under loss.  These issues, however, can be solved by developing \emph{QKD networks}. For a general survey of QKD, the reader is referred to \cite{QKD-survey-old,amer2021introduction,qkd-survey-new}.

\emph{QKD Networks} consist of multiple \emph{quantum repeaters} and \emph{trusted nodes} through which Alice and Bob may communicate.  These nodes allow for greatly improved distances between parties by allowing for shorter hops between nodes (thus less loss) and allowing for measurements to be performed between parties (in the case of trusted nodes).  Furthermore, a QKD network also has the added benefit that multiple paths may exist between Alice and Bob, allowing them to increase their overall key generation rates in an additive manner by distilling key-bits on each individual path connecting them.

To date, the majority of research in quantum networks spans two directions: the study of networks consisting of a majority of trusted nodes (with potentially some repeaters), and the study of networks consisting of only quantum repeaters (i.e., a true \emph{Quantum Internet} \cite{kimble2008quantum,caleffi2018quantum,wehner2018quantum}).  This represents the two extreme end-points of the timeline in quantum network development.  Quantum repeaters allow for two end-points to perform entanglement swapping, allowing them to establish Bell pairs even if the repeater is controlled by an adversary.  This provides a tremendously strong security guarantee, and also allows for the distillation of Bell pairs between parties which may be used for other applications including, and beyond, QKD (e.g., distributed quantum computing \cite{van2016path,yimsiriwattana2004distributed,cuomo2020towards} and distributed quantum sensing \cite{zhang2021distributed,PhysRevLett.121.043604,PhysRevLett.120.080501,PhysRevA.97.042337}).  Much current research is devoted to this future Quantum Internet, in particular to its behavior and performance in terms of maximizing entanglement distribution rates \cite{van2013path,azuma2016fundamental,hahn2019quantum,vardoyan2019stochastic,pirandola2016capacities,wallnofer2019multipartite,pirandola2019end,caleffi2017optimal,gyongyosi2017entanglement,pant2019routing,chakraborty2019distributed}, with \cite{caleffi2017optimal,gyongyosi2017entanglement,pant2019routing,chakraborty2019distributed} particularly focusing on routing within a Quantum Internet.

However, the technological requirements of a quantum repeater still surpass today's engineering capabilities, especially in terms of their need for short-term quantum memories.  Trusted nodes, on the other hand, are a technology available today and are currently used in all of the current large-scale metro area QKD networks (e.g., Vienna \cite{peev2009secoqc}, Hefie \cite{chen2010metropolitan}, Beijing-Shanghai \cite{zhang2018large}, and Tokyo \cite{sasaki2011field}).  This is due to the fact that trusted nodes simply behave like Alice or Bob and perform actual key distribution.  That is, they do not require any quantum storage but only classical storage.  While technologically feasible today, the downside, as their name implies, is that these nodes must be trusted as they will have full information on the final secret key.  Any corruption by an adversary on these nodes would jeopardize any secure communication in the network.  This security vulnerability, however, can be mitigated, to some extent, by taking advantage of the multiple paths generally afforded to users in a large-scale QKD network.  Much work has been devoted to studying routing strategies and general design strategies for QKD networks consisting of a majority of trusted nodes \cite{toliver2003experimental,le2007stochastic,wen2009multiple,tanizawa2016routing,yang2017qkd,han2014novel,mehic2019novel,yang2018quantum,li2020mathematical,yao2022efficient,tysowski2018engineering}.  For a general survey, the reader is referred to \cite{cao2022evolution,mehic2020quantum}.

In this work we consider a middle-ground, near-future, QKD network consisting of both trusted nodes and quantum repeaters, but with a majority of quantum repeaters in service, building off a network design we investigated in \cite{amer2020efficient}.  Such a system, though not as powerful as a true Quantum Internet, provides a realistic middle-ground between today's trusted node QKD networks and a future Quantum Internet.  The study of this QKD network can lay the foundation for a near-future secure quantum communication infrastructure as QKD technology is readily available commercially \cite{qkd-survey-new,amer2021introduction}; while the development of the technology for such a network can serve as a ``backbone'' for a future, true, Quantum Internet.  Furthermore, the theoretical tools and techniques that are developed to study and operate such a QKD network can lead to interesting insights in future quantum technologies.  Thus, though our network is specific to QKD applications, there is strong potential that the study of this network architecture can lead to important breakthroughs in quantum communication to be used when repeater and quantum memory technology are sufficiently advanced to build the true Quantum Internet.

In our previous work \cite{amer2020efficient}, we investigated two basic routing strategies and evaluated their relative performance under various network conditions.  In this work, we develop a new \emph{dynamic routing} strategy for this proposed near-future QKD network and show how it can dramatically improve overall secret key generation rates between two users.  
This dynamic routing strategy makes routing decisions based on the current network state---in particular based on the current size of relative raw-key pools between the various users and trusted nodes.  By altering the routing decisions in real time, we are able to significantly improve overall performance of the secret key generation rate as compared to standard ``static'' solutions, depending on the placement of the trusted nodes.  Beyond this, we also develop novel classical/quantum post-processing strategies, which further improve the overall performance of the QKD network, while also evaluating already known techniques in this network scenario.  In particular, we show how users can perform adaptive key-pooling based on the network topology, to improve the overall key-rate of the system.

Taken together, we make several contributions in this paper.  To our knowledge, we are the first to propose dynamic routing strategies for QKD networks consisting of a majority of quantum repeaters (past work in dynamic routing has focused only on trusted-node networks \cite{yang2017qkd,yang2018quantum,chen2021ada} where the routing challenges are complementary, but different).  The dynamic routing strategy we propose (both the actual strategy and the lessons learned from its development) may be highly beneficial to a future, true, Quantum Internet (e.g., by adapting the entanglement generation paths based on observed path depolarization noise).  We discuss several cases where our strategy greatly outperforms static solutions, while also showing when dynamic routing may not be as helpful, thus leading to guidelines for future practitioners of this network technology.  Second, we propose novel classical post-processing strategies involving the dynamic pooling of raw key bits into distinct blocks based on network characteristics.  We also employ \emph{Classical Advantage Distillation} (CAD) \cite{CAD}, adapting the CAD parameters appropriately for each distinct pool.  Our processing strategies developed in this work are actually broadly relevant to other QKD systems (i.e., not only to users running QKD protocols in our near-future network) and may even be used to improve current-day systems.  Finally, we perform a rigorous set of evaluations of our proposed routing and post-processing strategies under a variety of network scenarios; in particular, we consider network size, node quality, link quality, and resource placement within the network.  Our methods and approaches may be widely applicable to future Quantum Internet solutions and also to current-day QKD systems (even running on a single point-to-point link).





\section{Network Model}\label{sec:model}

In this work, we study a near-future, large-scale, QKD network as in our recent paper  \cite{amer2020efficient}, itself inspired by standard entanglement based quantum networks~\cite{pant2019routing}.  This network is modeled as a graph with nodes representing either an end-user (Alice or Bob who wish to establish a shared secret key), a quantum repeater, or a trusted node.  End-users and trusted nodes will be given labels $T_0, T_1, \cdots, T_{n+1}$ with $T_0$ being Alice, $T_{n+1}$ being Bob, and all other $T_i$ being trusted node units (thus there are $n\ge 0$ trusted nodes in the network).  Each pair $T_i$ and $T_j$ have a \emph{raw key pool}, denoted $\mathcal{RK}_{i,j}$, and a \emph{secret key pool}, denoted $\mathcal{SK}_{i,j}$, storing shared key material between those nodes.  The raw key consists of those shared key bits \emph{before} error correction and privacy amplification are run, whereas the secret key consists of those secret key bits stored following the completion of these classical processes.

Each edge in this graph connecting two nodes represents  a fiber link allowing for the transmission of qubits between nodes; this link may be noisy and lossy.   We assume that all nodes have a quantum memory capable of storing a single qubit for every neighbor they are attached to, but this qubit can be stored only for a short amount of time (in particular, the qubit can only be stored for a single network round, to be discussed), after which it must be discarded.  The memory may also be noisy causing the state to decohere with a certain probability.  Later, when evaluating our routing strategies, we will consider a grid topology with the end-users near opposite corners. However, nothing in this section is specific to this one type of topology.  For a grid topology, all nodes, except those on the outer border, have four neighbors and are capable of storing at most one qubit per neighbor.

Our network operates over multiple \emph{rounds}, each round being divided into three distinct \emph{stages}: \emph{Entanglement Distribution}, \emph{Entanglement Routing}, and finally \emph{Key Distribution}.  A fourth and final stage \emph{Secret Key Establishment and Routing} is performed at the conclusion of the network simulator (though, in practice, can be performed periodically as key pools are established).

{\bf Stage 1.} In the first stage, each pair of connected nodes $u$, $v$ in the network (which, again, may represent an end-user, a repeater, or a trusted node) attempts to share half of an entangled pair with each other over the fiber channel.  This process succeeds with probability $P=10^{-\alpha*L/10}$, where $L$ is the length of the fiber connecting $u$ and $v$ and $\alpha$ is a constant. Note that as the distance between nodes increases, the loss also increases exponentially.  If this is unsuccessful, the qubit is lost and no shared entanglement exists between these nodes for this round of the network. In addition, we also consider that a channel may cause the entanglement to decohere with some probability $D$ (this may be due to noise in the fiber channel, noise in the memory system, or both combined). Thus, at the end of this stage, with probability $P$, each pair of neighbors $u,v$ share the state 
\begin{equation}\label{eq:ms-pair-state}
  \rho_{u,v} = (1-D)\ket{\Phi^+}\bra{\Phi^+} + D\frac{I}{4}
\end{equation}
where $\ket{\Phi^+} = \frac{1}{\sqrt{2}}\left(\ket{00}+\ket{11}\right)$ and $I$ is the identity matrix. Otherwise, with probability $1-P$, those neighbors share the vacuum state (i.e., they share no state).  We assume nodes are able to determine whether they have a vacuum or not; of course they cannot determine if the state has decohered.  Note that, as mentioned earlier, we assume nodes are capable of storing entangled pairs in a short-term memory that remains coherent only for a single network round (after which, any qubit stored in memory is lost).

{\bf Stage 2.} Following the first stage, the second stage of the network round begins.  Here, a routing algorithm is executed (to be discussed in Section~\ref{sec:routing}), which must decide how to effectively route entanglement between nodes $T_i$ and $T_j$ (i.e., between end-users and trusted nodes).  This routing algorithm can take into account information including and beyond: the state of the memory at any node $u$ (i.e., whether there is or is not a qubit in $u$'s memory with neighbor $v$); past history of link establishment (e.g., whether a link is very noisy); current status of the key-pools between any $T_i$ and $T_j$; expected key generation rates between nodes $T_i$ and $T_j$; and network topology, including trusted node locations.  However, this routing algorithm cannot take into account whether any qubit pair shared between nodes $u$ and $v$ is a proper Bell state or a decohered one. 

%

Note that, unlike routing in entanglement networks, here we have several interesting optimization decisions unique to our scenario.  Primarily, the routing algorithm's goal in our QKD network is to maximize the \emph{secret} key generation rate between $T_0$ and $T_{n+1}$.  The algorithm may take advantage of the existence of trusted nodes operating between the end-users.  Due to the effects of depolarization noise (which amplifies over path distance), maximizing the number of shared states between any $T_i$ and $T_j$ does not necessarily lead to maximal secret key generation rates as we will show later.  Thus, new methods are required to fully harness the capabilities of this network.

Given the information available,
the routing algorithm will determine a set of paths, each path starting and ending with a node of the form $T_i$ and $T_j$ and with only quantum repeaters as interior nodes in these paths.  That is, these paths are from end-users/trusted nodes to other end-users/trusted nodes and contain only quantum repeater nodes in the middle.  These paths will be used by the repeaters to attempt to create an entangled pair between users $T_i$ and $T_j$ by performing entanglement swapping operations at each repeater in the path. If successful, this will create a ``virtual path'' (i.e., a shared entangled state) between nodes $T_i$ and $T_j$ by using the states shared between respective repeaters.  Because of this, the paths must be disjoint in their use of the available stored qubits (i.e., two different paths cannot rely on a single qubit).

In more detail, given a path $T_i \rightarrow u_1 \rightarrow \cdots \rightarrow u_k \rightarrow T_j$ consisting of $k$ repeaters, each repeater, $u_i$, will perform a Bell state measurement on their respective half of the shared state $\rho_{u_i, u_{i+1}}$,  see Equation (\ref{eq:ms-pair-state}). This succeeds, independently for each repeater, with probability $R$.   Since we assume each of the repeaters acts simultaneously, if a single repeater on this path fails, the entire path must be discarded.  Note that, since the state shared between any pair on this path also may be completely mixed, following this stage there are two outcomes: with probability $R^k$, nodes $T_i$ and $T_j$ share a state of the form:
\begin{equation}\label{eq:ms-final-state}
  \rho_{T_i, T_j} = (1-D)^{k+1}\kb{\Phi^+} + [1-(1-D)^{k+1}]\frac{I}{4}.
\end{equation}
Otherwise, with probability $1-R^k$, nodes $T_i$ and $T_j$ do not share a quantum state from this path and all nodes discard any qubits that would have been used for this path.  As in the first stage, we assume nodes $T_i$ and $T_j$ will know whether this path entanglement attempt was successful or not; however, they do not know if the state is the desired Bell state $\ket{\Phi^+}$ or a completely mixed state.  Observe that the decoherence probability increases with path length.

{\bf Stage 3.} In the third stage, with entanglement shared, each pair $T_i$ and $T_j$ can conduct the quantum portion of their chosen entanglement based QKD protocol, in our case the E91 QKD protocol \cite{QKD-E91}. If the two parties select the same randomly chosen basis, which occurs with probability $1/2$, the two parties add the resultant raw key bit to their raw key pool $\mathcal{RK}_{i,j}$. Otherwise, no addition to their raw key pools is made.  At the end of this stage, all remaining qubits in the system are discarded, as we are assuming quantum memories can only store qubits for a single network round.  The network then repeats these three stages for the next round.

{\bf Stage 4.} When the network has been running for a sufficiently long period of time, the fourth stage is executed.  Here, error correction and privacy amplification are run between nodes $T_i$ and $T_j$ adding secret key material to $\mathcal{SK}_{i,j}$.  As we are interested only in the asymptotic performance of our routing strategies, we assume perfect error correction and use asymptotic key rates for the E91 protocol.  In particular, this means if the error rate in $\mathcal{RK}_{i,j}$ is $Q_{i,j}$, then $|\mathcal{SK}_{i,j}| = |\mathcal{RK}_{i,j}|(1-2h(Q_{i,j}))$ \cite{QKD-BB84-rate1,QKD-renner-keyrate} where $h(x)$ is the binary entropy function.

{\bf Final Stage.} Finally, as the ultimate goal of the network, of course, is for Alice and Bob to share as much secret key material as possible, this last stage of the network is to push secret key material from each $\mathcal{SK}_{i,j}$ to $\mathcal{SK}_{0,n+1}$ (i.e., the secret key pools of Alice and Bob). For this, we construct a flow network consisting of Alice, Bob, and the trusted nodes, where an edge with capacity $c$ exists between parties $i,j$ if $|\mathcal{SK}_{i,j}| = c$. We then use standard flow network techniques to find a maximal flow of secret key material from Alice to Bob over the trusted nodes. Here, flowing $x$ bits from party $i$ through party $j$ to party $k$ signifies that trusted node $j$ publishes the XOR of the first $x$ bits of $\mathcal{SK}_{i,j}$ and $\mathcal{SK}_{j,k}$, allowing party $k$ to recover the value $\mathcal{SK}_{i,j}$, removing those $x$ key bits from $\mathcal{SK}_{i,j}$ and appending them to $\mathcal{SK}_{j,k}$. At the end of this, we will have succeeded in pushing the maximal amount of key material to Alice and Bob, maximizing $|\mathcal{SK}_{0,n+1}|$. We note that if the flow is not saturating, it is possible that there remain some parties $i,j$ for which $|\mathcal{SK}_{i,j}| > 0$ causing waste (as some trusted nodes have left-over secret key material that cannot be pushed to the end-users Alice and Bob).  One of our main contributions in this work is to develop a {\em dynamic} routing strategy to balance network operation to identify and avoid such inefficiencies.

While, in practice, this last stage would be run periodically during the network's operation, in this work we are only interested in the asymptotic behavior of our routing solutions and, so, we run this fourth stage at the very end of our simulations.  When designing routing strategies, then, our main metric for consideration is the final secret key generation rate between Alice and Bob, namely the ratio $|\mathcal{SK}_{0,n+1}| / N$ where $N$ is the total number of network rounds simulated.  The overall goal of our network and routing algorithms will be to maximize this key-rate ratio.




\section{Dynamic Routing to Maximize Key Rate} \label{sec:routing}

In this section, we first motivate the need for dynamic routing, and then describe our dynamic routing strategy.  


\subsection{The Need for Dynamic Routing} \label{sec:need-dynamic}
As background, we first briefly describe a 
routing algorithm for a QKD network with both quantum repeaters and TNs~\cite{amer2020efficient}. 
This algorithm assumes that link-level entanglement state, i.e., whether the entanglement between two neighbor nodes has been successfully established or not in Stage 1, is known to all $\{T_0,\ldots,T_{n+1}\}$ in the network. This information can be communicated using classical channels, and is realistic in small-scale networks. The routing algorithm 
works by selecting 
the shortest path (in number of hops) between any pair of nodes in $\{T_0,\cdots,T_{n+1}\}$, where recall that $T_0$ represents Alice and $T_{n+1}$ represents Bob, and  $T_1,\ldots,T_n$ are TNs. When two paths of equal length are found, one of them
is selected randomly. Once a path between $T_i$ and $T_j$ is found, then all the links along the path are removed (since qubits can only be used once in a path), and the procedure repeats for the remaining links until no path can be found.

Intuitively, the above algorithm may be effective since shorter paths 
tend to have lower decoherence noise and higher overall Bell statement measurement (BSM) success rate, and hence are more likely to succeed in establishing keys. The study in~\cite{amer2020efficient} demonstrates its effectiveness. This algorithm, however, follows a static strategy, i.e., always favoring paths with shorter distances. As alluded earlier, such static strategies may lead to waste in keying materials, for instance, when there are asymmetries in the placement of TNs relative to Alice and Bob.

\begin{figure}
    \centering
    \includegraphics[width=.35\textwidth]{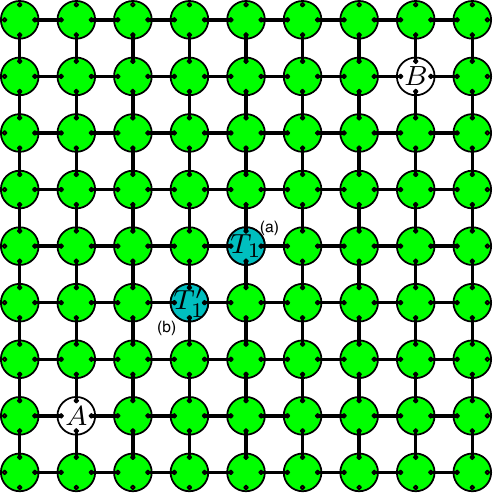}
    \caption{An example illustrating the need for dynamic routing. }
    %
    \label{fig:central_and_dev_configs}
\end{figure}

One example is shown in Fig \ref{fig:central_and_dev_configs}, where Alice and Bob are along the diagonal of a grid network, and one TN is placed in the network. We consider two settings: (i) a single TN, $T_1$, is placed in the center of the grid network, equal distance to Alice and Bob, and (ii) a single TN, $T_1'$, is placed along the diagonal, but closer to Alice than Bob. We refer to the first setting as an ideal setting since the TN is of equal distance to Alice and Bob, and refer to the second setting as the \textsc{off-center} setting.  
{With BSM success rates of 0.85, decoherence rates of 0.02, and link lengths of 1 km, we observe that the key rate achieved using the above static routing algorithm in the \textsc{off-center} setting is a 76\% decrease from that of the ideal setting (the key rate is 0.037 versus 0.153). This is because in the \textsc{off-center} setting, the shorter distance between Alice and $T_1'$ causes key establishment between them to be placed at a higher priority, leading to significantly more secret key bits between Alice and $T_1'$ than those between $T_1'$ and Bob, which end up being wasted since the final key rate between Alice and Bob requires keys between both Alice and $T_1'$, as well as between $T_1'$ and Bob. } 


\subsection{Dynamic Routing Algorithm}
To resolve the wastage issue described above, we design a dynamic routing strategy that identifies which key pools should be prioritized 
to properly balance the key-pools in the network in order to maximize the key rate 
between Alice and Bob. 
In the following, we first describe how we identify the proper key pools to prioritize and then how we balance the key pools. 


Consider graph $\mathcal{G}$ with the set of nodes $\{T_0,\ldots,T_{n+1}\}$ and the edges $(T_i,T_j)$, where $T_0$ is Alice ($A$) and $T_{n+1}$ is Bob ($B$). Let $c(T_i,T_j)$ be the capacity of edge $(T_i,T_j)$, which  represents the amount of secret key bits that have been created between these two nodes $T_i$ and $T_j$. Note that this quantity can be easily estimated by analyzing which paths were used to generate the raw key bits between $T_i$ and $T_j$ and 
the properties of these paths (see Section~\ref{sec:model}), even if the raw key bits shared between $T_i$ and $T_j$ have not been actually converted to secret key bits. 

A natural approach to achieve more balanced keys 
is finding low-capacity edges $(T_i,T_j)$ and then allowing the nodes on the two ends of the edge (i.e., $T_i$ and $T_j$) to have a higher priority to establish paths and hence keys. This approach, however, may not improve the key rate between Alice and Bob. This is because the low key rate between $T_i$ and $T_j$ may be because they are far away from each other; even if they are close to each other, creating more key bits between them may not help Alice and Bob, 
since what is needed is actually more key bits between Alice and Bob.


In the following, we present a novel 
approach that finds edges to prioritize. The main idea is that we first locate the edges in the network that have surplus key  material (i.e., more than the key materials on the other edges), and then determine how best to spread the network resource usage away from those edges in upcoming iterations. Specifically, in each iteration, we first identify the edge with the highest capacity (i.e., the maximum amount of secret key bits) in our flow network $\mathcal{G}$. Let $e_{\max} = (T_i,T_j)$ denote this edge and  $c_{\max}$ denote its  capacity. We next locate the shortest 
$A \rightarrow T_i$ and $T_j \rightarrow B$ 
paths in the flow network $\mathcal{G}$, so that $A \rightarrow T_i-T_j \rightarrow B$ is the shortest path 
from Alice to Bob that also includes edge $(T_i,T_j)$. In the above,  there may exist other TNs along path $A \rightarrow T_i$ or $T_j \rightarrow B$. (Note that for ease of exposition, we assume a single shortest path $A \rightarrow T_i-T_j \rightarrow B$; the case with multiple shortest paths can be treated similarly as follows.)  
Along path $A \rightarrow T_i-T_j \rightarrow B$, we identify the edges that are \textit{under-full} with respect to $e_{\max}$, i.e., any $e_i$ on the path such that $(1+\sigma)c_i \leq c_{\max}$, where $\sigma \ge 0 $ is a parameter chosen beforehand. Let $E_{\text{uf}}$ denote the set of such under-full edges. Let $e_{\min}$ be the edge in $E_{\text{uf}}$ such that its capacity is the lowest, and let $c_{\min}$ denote its capacity. We then prioritize any edges $e_i \in E_{\text{uf}}$ so that $c_i \leq (1+\delta)c_{\min}$ for some tolerance value $\delta \ge 0$ that is determined beforehand. Let $E_{p}$ denote this set of edges to be prioritized. Since it is not beneficial to establish keys between two TNs that are far away from each other, we remove from $E_{p}$ all the edges $(T_k, T_k')$ with $dist(T_k, T_k') \geq \theta \cdot dist(A,B)$, where $dist(\cdot)$ is the distance between two nodes, and $\theta \in (0,1)$ is a parameter that we set to be $0.75$ in the rest of the paper. 

Once $E_{p}$ is identified, we simply modify the static global  routing algorithm (see Section~\ref{sec:need-dynamic}) as follows. Rather than choose the shortest paths among all pairs of nodes, we first prioritize the shortest paths among the pairs of nodes specified by $E_{p}$. Once all such paths have been attempted, and there are no more paths between the prioritized pairs of nodes, we fall back to the standard operation, attempting all remaining node pairs ordered by distance.

In the special case where there is a single TN $T_1$ between Alice and Bob, the above algorithm basically reduces to the following. It first identifies one edge, $(A,T_1)$ or $(T_1,B)$, as $e_{\max}$, and the other edge as $e_{\min}$. If the capacities of these two edges differ significantly, specifically, $(1+\sigma) c_{\min} \le c_{\max}$, it then prioritizes $e_{\min}$ until in one iteration the capacity of $e_{\min}$ is close to that of $e_{\max}$, specifically, when $(1+\sigma) c_{\min} > c_{\max}$. 

\begin{figure}[t]
	\centering
    \subfigure[Topology.]{%
		\includegraphics[width=0.31\textwidth]{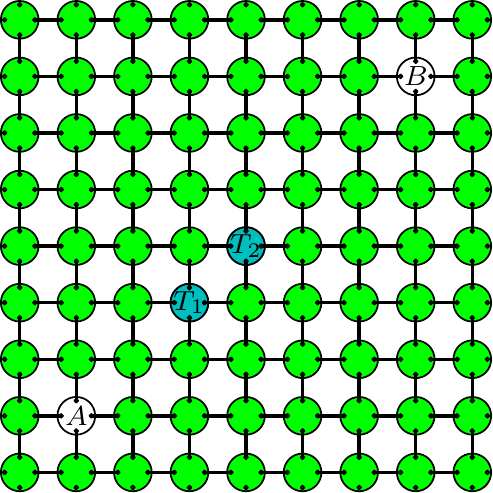} 
		\label{fig:uneqdiag}
    }
    \hspace{.05in}
	\subfigure[Example scenario.]{%
		\includegraphics[width=0.31\textwidth]{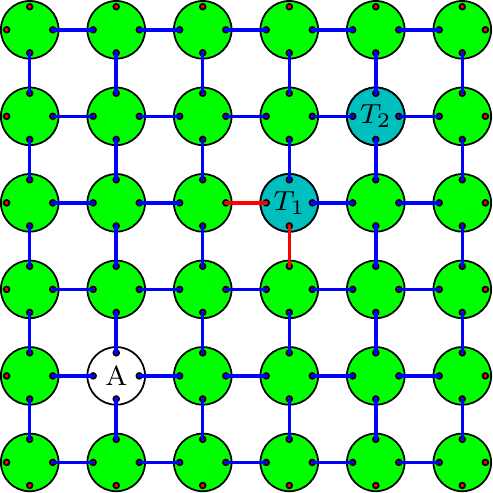}
		\label{fig:balex1}
	}
    \hspace{.05in}
	\subfigure[Static vs. dynamic routing.]{%
		\includegraphics[width=0.31\textwidth]{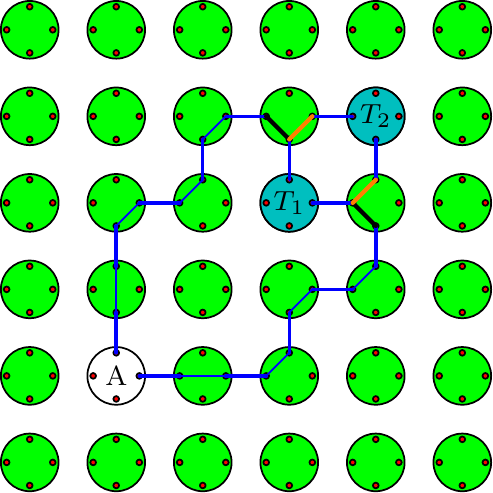}
	    \label{fig:balex2}
	} 	
	\caption{An example illustrating the difference in routing decisions by the static and dynamic routing strategies.}
    \label{fig:2TN-ex}
\end{figure}

We next illustrate the above dynamic routing algorithm using an example with two TNs, as shown in Fig.~\ref{fig:2TN-ex}. 
Fig.~\ref{fig:2TN-ex}a shows the topology, which has two TNs, both along the diagonal of the grid topology, one ($T_1$) is closer to $A$ and the other is equal distance to $A$ and $B$. 
Suppose that in one iteration, the two red links incident to $T_1$ in Fig.~\ref{fig:2TN-ex}b were not established successfully (i.e., entanglement generation fails), and the rest of the links were established successfully. In this scenario, Fig.~\ref{fig:2TN-ex}c illustrates the different decisions of static routing vs. dynamic routing. In static routing, the remaining two links incident to $T_1$ will be used to connect $T_1$ and $T_2$ along two paths, each of two hops, requiring the entanglement swapping operation highlighted in orange. In contrast, our dynamic routing strategy will use those edges as part of a length six path between $A$ and $T_1$ by making the internal connection highlighted in black in Fig. \ref{fig:2TN-ex}c. 

\begin{figure}
    \centering
    \includegraphics[width=\textwidth]{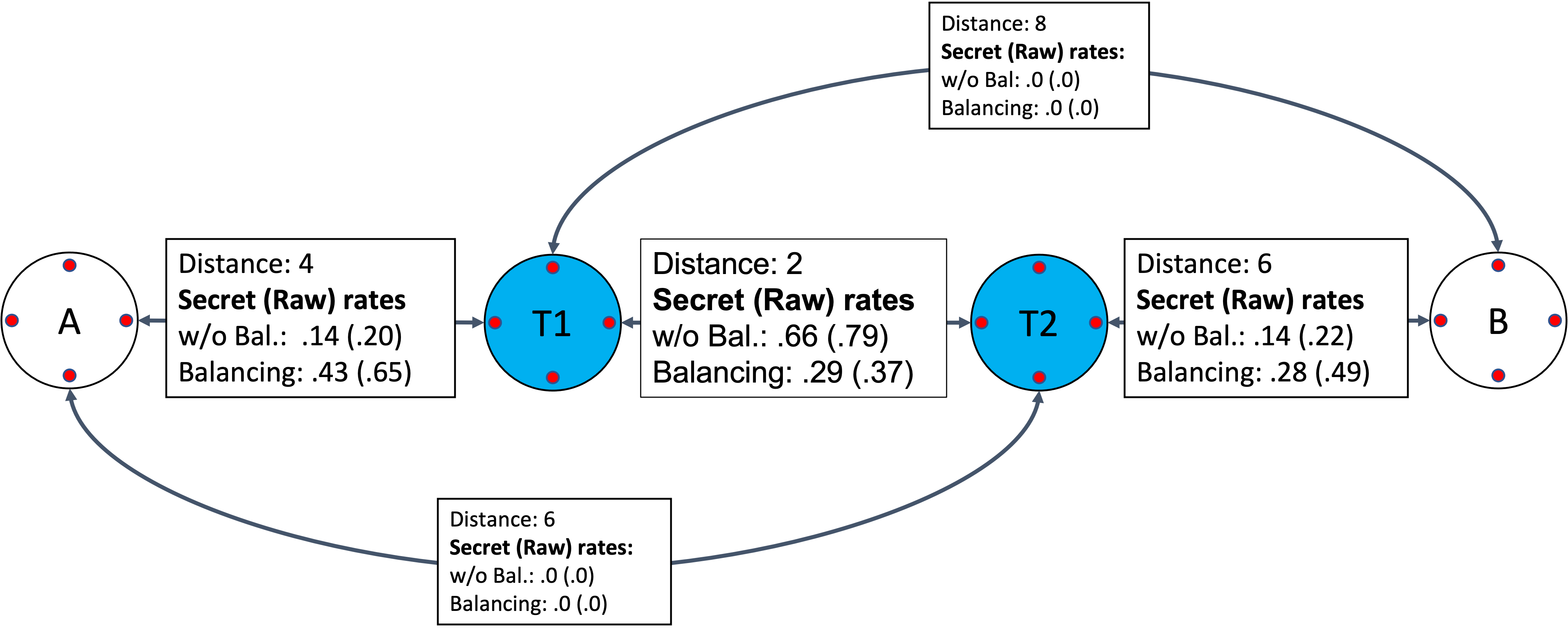}
    \caption{Illustration of the benefits of the dynamic routing strategy: comparison of the key rates on various edges of the network flow graph when using static and dynamic routing strategies. 
    }
    \label{fig:balex3}
\end{figure}

Fig. \ref{fig:balex3} shows the network flow graph for the setting in Fig. \ref{fig:2TN-ex}a. It has  five edges, $(A,T_1)$, $(A,T_2)$, $(B,T_1)$, $(B,T_2)$, and $(T_1,T_2)$, with Manhattan distance 4, 6, 8, 6, and 2, respectively; the edge $(A,B)$ is not included in the graph for clarity. 
For static routing, the two edges, $(A,T_1)$ and $(T_2, B)$, each have a secret key rate of 0.14, while the $(T_1, T_2)$ edge enjoys a much higher key rate of 0.66. When using our dynamic strategy, we sacrifice some key rate between $T_1$ and $T_2$ (dropping it to 0.29) in order to augment the key rate of edges $(A,T_1)$ and $(T_2, B)$, leading to an overall key rate of 0.28 between Alice and Bob, twice as high as the key rate (0.14) under static routing.

\section{Classical Post-Processing}\label{sec:adaptive-routing}

 
The dynamic routing strategy as described earlier, while reducing the wastage in the network, can also lead to long paths and paths with more diverse lengths, which can be detrimental to the final secret key distillation.  
We next consider two classical post-processing methods to address the above issues. 
The first method, {\em key-pool segmenting}, is a novel method to address the heterogeneous noise rates  that can be caused by diverse path lengths.   
The second method, classical advantage distillation (CAD) \cite{BB84-CAD}, is a well-established concept often applied to BB84 and E91, though we are not aware of any research evaluating its performance in multi-path QKD networks.  CAD sacrifices raw key material in order to reduce the noise rate and increase the overall secret key rate; it is particularly beneficial in point to point QKD systems with high noise channels allowing BB84 to surpass the usual $11\%$ noise tolerance \cite{BB84-CAD,chau2002practical,CAD-3}. Both processes, our key-pool segmenting strategy and also CAD, can be beneficial alone, and, as we shall show, combining them can lead to even larger benefits. 

\subsection{Key-pool Segmenting}
In standard QKD protocols, one usually assumes that the quantum channel used to transmit quantum states between two parties has a certain fidelity, and
each raw key bit shared between these two parties has the same error probability.
In our network, there are multiple paths that could be used to share entanglement between $T_i$ and $T_j$, which may not be of the same length. Therefore, even if each individual link in the network has the same fidelity, these paths may result in entanglement with different fidelity.
Specifically, in one iteration, the shortest path may be used between $T_i$ and $T_j$, resulting in a key bit with the lowest possible noise; in another iteration, maybe only a longer, noisier path can be used, resulting in a noisier raw key bit. The key-pool between $T_i$ and $T_j$ therefore is a mixture of key bits with different expected noise rates, with the expected noise of the entire key-pool being a weighted average thereof.

This mixture of key-pools can degrade the overall quality of the key-pool to the point that the total key rate is reduced (see example below).  To mitigate this impact, we propose a segmenting approach to post-processing: rather than process the entire raw key together, we allow the network to segment the raw key-pool based on the expected noise rate of each bit, partitioning it into sub-pools with identical expected noise rates (or, if necessary, noise rates within a given range). When two parties wish to transform their raw keys into secret-keys, they process each sub-pool separately, before ultimately recombining it into a single key-pool. This process can result in substantial increases in key-rate in certain network scenarios. 
Furthermore, it is practical for the network to realize, as the repeaters can periodically disclose (through classical transmission channels) their previous routing decisions to allow parties to discover which paths were used. The length of these paths, assuming relatively static noise profiles on the links or periodic channel tomography, can then be used to find the expected noise rate of each bit.  Unlike point to point systems, our segmenting strategy is particularly suited to large scale QKD networks such as this, since one can be certain of the relative noise of different paths and this can be estimated by disclosing routing decisions.

We give an example here, and then generalize it to show that, in the asymptotic regime, this segmenting approach results in strictly better or equal key-rates than processing the key-pool as a mixture. Consider a raw key composed of a mixture of two sub-pools, $K_1$ and $K_2$, where $K_1$ has an expected noise rate of $Q_1 = 0.04$ and compromises $p_1 = 75\%$ of the total key-pool, and $K_2$ has an expected noise rate of $Q_2 = 0.10$ and compromises the remaining $p_2 = 25\%$ of the raw key. The total average noise-rate is 
$Q_{avg} = 0.055$. 
Using the standard BB84 key-rate equation, this corresponds to a key-rate of $1-2h(Q_{avg}) = 0.385$. The segmented key-rate is simply the weighted average of the key-rate of each sub-pool, resulting in a key-rate of $p_1[1-2h(Q_1)] + p_2[1-2h(Q_2)] = 0.402$, leading to a $4.4\%$ increase in key rate compared with the naive approach. 

In general, this segmented key-pool approach will never lead to a lower key-rate, and can only improve the asymptotic key-rate when there are at least two key pools with different noise levels.  This is easily seen due to the concavity of Shannon entropy on the interval $[0,1/2]$.  In the finite key-length regime, there may be sampling considerations that do not allow segmenting to guarantee equal or better performance. We leave the finite key length case as future work.

\subsection{Classical Advantage Distillation (CAD)}
After key-pool segmenting as described earlier, some key pools may have high noise. For such pools, 
classical advantage distillation (CAD) \cite{CAD} is a well-studied technique that may be used to extract additional secret key material from noisy raw key-pools.
We next briefly describe CAD. In Section~\ref{sec:results}, we evaluate the performance of CAD with key-pool segmenting and dynamic routing.

CAD, like segmenting, is an additional  post-processing method undertaken by two parties once their raw key-pool has been generated. For our analysis, we use specifically the CAD process discussed in \cite{BB84-CAD,chau2002practical}. Prior to error correction, the two parties engage in the following procedure, sacrificing some raw key material in order to achieve an advantage on the remaining key-material. For ease of exposition, we describe CAD assuming the two parties are  Alice and Bob; in our setting, they can be any two parties, $T_i$ and $T_j$.  For some CAD level $C \geq 1$:
\begin{enumerate}
    \item Alice selects $C$ raw-key bits that all have the same value, sending the $C$ corresponding indices to Bob.
    \item Bob notifies Alice whether or not, in his raw key,  those $C$ bits take the same value.
    \item If these $C$ bits are consistent in Bob's key (they do not need to be consistent with Alice's key), then Alice and Bob shrink the block into a single bit, and add the bit to a new raw key; otherwise they discard the $C$ bits. 
    \item They repeat this process until the entire original key has been processed.
\end{enumerate}
Let $CAD[C]$ represent  running CAD with blocks of size $C$. It is clear that after $CAD[C]$, the raw key will shrink to, at a minimum, $1/C$ of the original length.
%
If the key is noisy, it may shrink by an even greater amount, as there will be a greater number of blocks discarded by Bob. At low noise levels, CAD does not result in an increase in key-rate (or equivalently, $C=1$ is optimal), but as the noise level increases, the optimal $C$ also increases. For $C=2$, the noise tolerance of the 4-state E91 protocol increases from the usual $11\%$ to $18.2\%$ \cite{BB84-CAD,chau2002practical}. In our work, we determine the key rate for E91 with CAD level $C$ using the work in \cite{BB84-CAD}, where it was found to be: 
%
\begin{align*}
r(C) = 1-h(\alpha) - (1-\alpha)h\left(\frac{1-\beta^C}{2}\right)
    - \alpha h\left(\frac{1-\gamma^C}{2}\right),
\end{align*}

where
$$\alpha = \frac{Q^C}{Q^C+(1-Q)^C},\quad \beta = \frac{1-3Q+2\lambda}{1-Q}, \quad  \gamma = \frac{|Q-2\lambda|}{Q},$$
and $Q$ is the noise rate in the raw key, and $\lambda$ is optimized in $[0,Q].$

Applying CAD to our network can result in increases in key-rate for higher noise scenarios. This is true both with and without our aforementioned key-pool segmenting technique. When the key-pool segmenting technique is used, we utilize the noise estimates for each individual sub-pool to select the optimal CAD level. In practice, this estimate can be obtained based on the network characteristics. 

While CAD is a well-established tool, we discuss it here for two reasons. The first is to apply it to our networks and quantify any resulting benefit. The second is to demonstrate that in our contexts, there is an interesting interplay between CAD and the key-pool segmenting technique that we propose: in some network configurations, combining them can lead to positive key rates, which cannot be achieved using one technique alone.


\mysection{Evaluation Results} \label{sec:results}


We use simulations to evaluate the performance of the dynamic routing strategy, as well as the impact of key-pool segmenting and CAD on the resultant key rate.  In all of our simulations, we use a custom-built simulator written in Python. 
All simulations were conducted over $10^6$ network rounds, or iterations. 
The networks we investigate here are square lattices of size $S$, with Alice and Bob located at the corners, embedded in an another square lattice of size $S+2$. 
We limit our analysis to consider the decoherence rate as our independent variable, as our previous results indicate that this rate is the primary factor determining the performance of the network.  
The rest of the parameters include the lattice size ($S = 7$), the length of the fiber connecting two neighboring nodes ($L=1$ km), the probability of
success in the fiber ($P=10^{-.15L/10}$), and Bell State measurement success probability ($R=0.85$). 
Note that while our evaluation focuses on this grid setting, our dynamic routing strategy and post-processing techniques can be applied to general topologies. 

\subsection{Results of Dynamic Routing}\label{sec:EvalDynam}
We first analyze the performance of our dynamic routing strategy in several scenarios with asymmetric TN placement. For each case, we present the performance of our dynamic routing algorithms as decoherence increases, and compare the key rates to those that can be achieved using ideal TN placements. For the dynamic routing strategy, we set the tolerance parameter $\sigma = 0.15$ as a reasonably tolerant threshold, as well as $\delta = 0.05$, although there may be potential for optimization on different topologies and configurations. 

\dubfigsingle{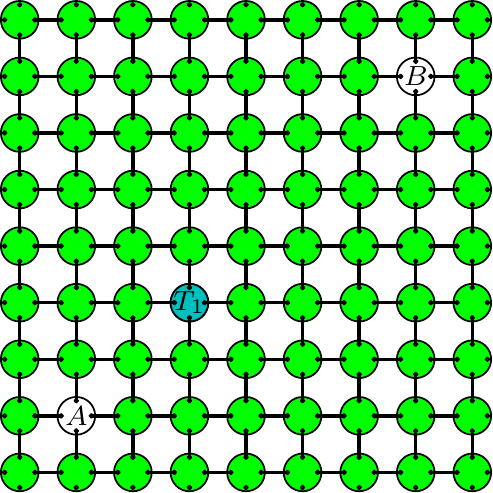}{\textsc{off-center} setting.}{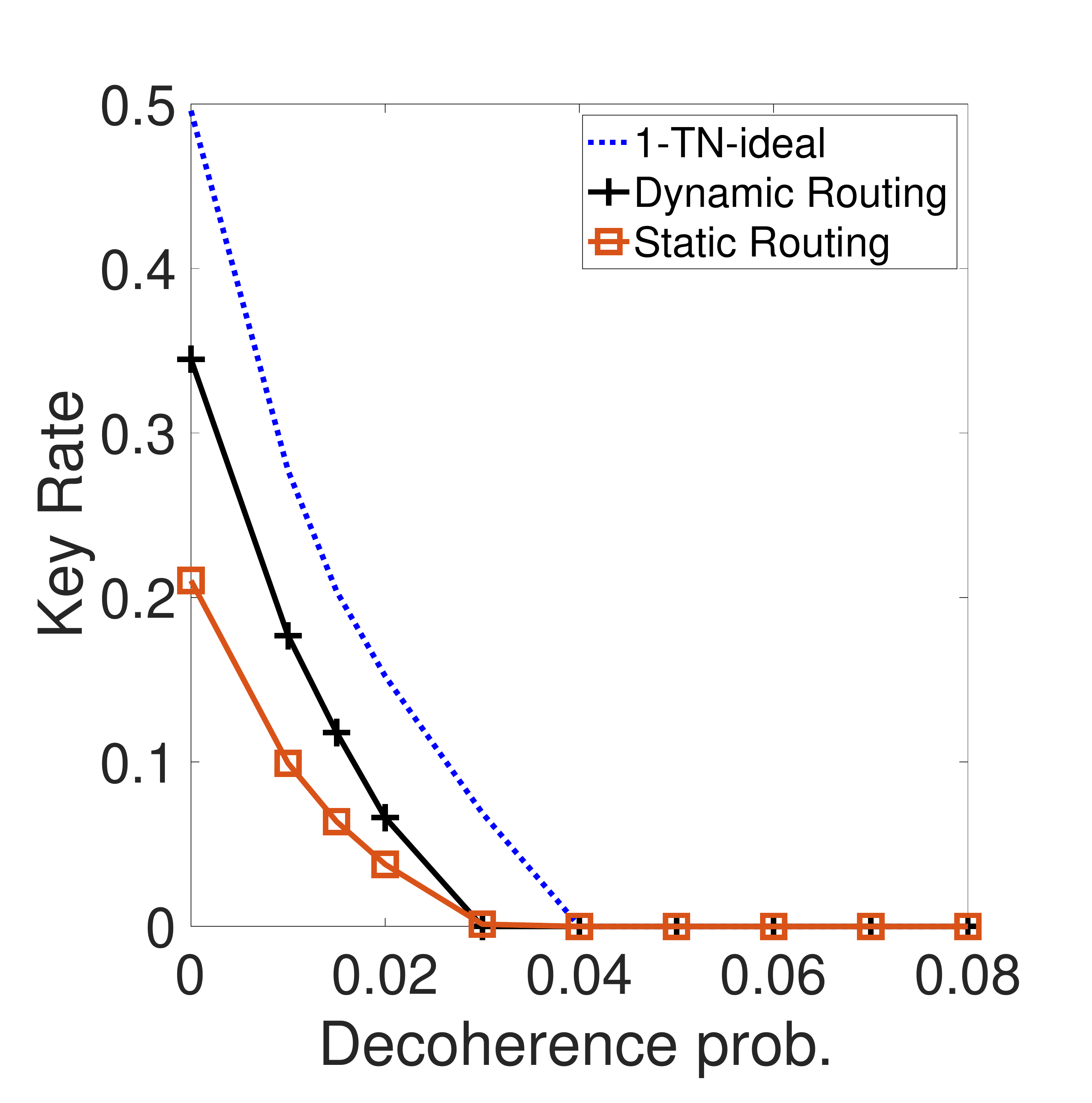}{Key rate.}{Performance of dynamic routing with a single TN that is not placed in the center. 
}{OffCenter} 

\smallskip
\noindent{\bf Single TN.} We first consider a setting with a single TN that is not placed in the center of the grid (see Fig. \ref{fig:OffCenter}a). 
Specifically, the TN is placed along the diagonal, closer to Alice than Bob, referred to as \textsc{off-center} setting. 
%
Fig. \ref{fig:OffCenter}b shows the key rate under static and dynamic routing. 
We see that dynamic routing leads to significantly higher key rate than static routing. For instance, when there is no decoherence, dynamic routing leads to a key rate of 0.345, 64\% higher than the key rate under static routing (which is 0.210). When the decoherence probability is 0.02, dynamic routing leads to a key rate of 0.066, 78\% higher key rate than that under static routing (0.037). 
For reference, Fig. \ref{fig:OffCenter}b also plot the results when the TN is placed at an ideal location (i.e., in the center of the grid,  equal distance to Alice and Bob). 
As expected, even under dynamic routing, the key rate in \textsc{off-center} TN placement is lower than that under the ideal placement, due to the longer distance between the TN and Bob. 
%
%



\begin{figure}[ht]
	\centering
	\subfigure[\textsc{diag-2-6-4} setting.]{%
		\includegraphics[width=0.30\textwidth]{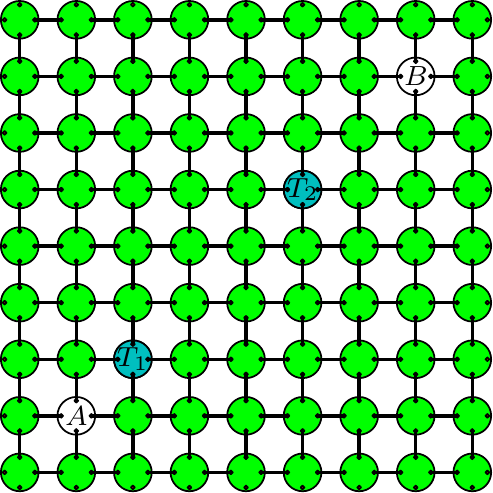}
		\label{fig:uneqdiag1}
	}
	\subfigure[Results for \textsc{diag-2-6-4} setting.]{%
		\includegraphics[width=0.30\textwidth]{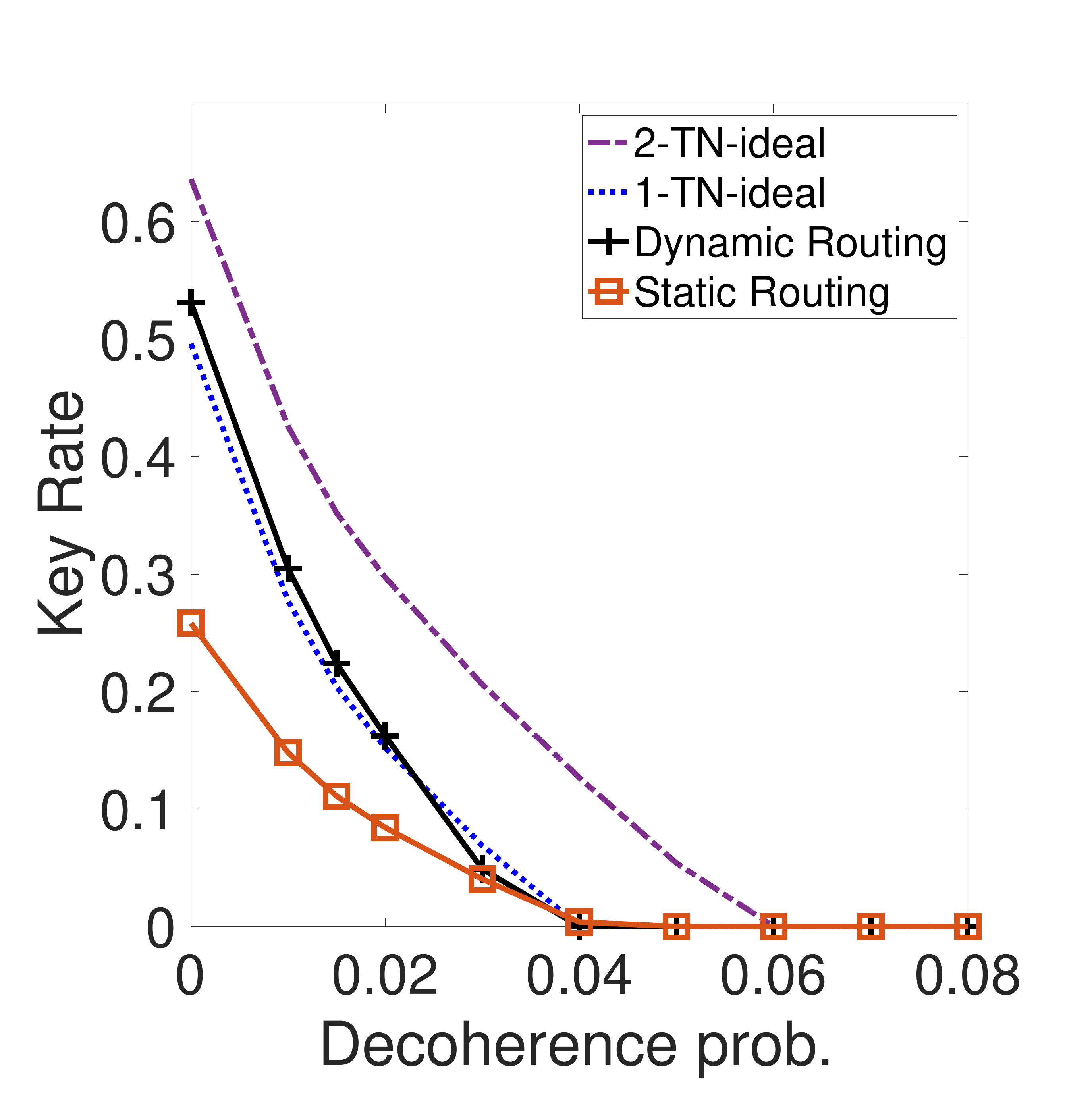}
		\label{fig:UneqDiagGlobal1}
	}\\
    \vspace{-0.1in}
	\subfigure[\textsc{diag-4-2-6} setting.]{%
		\includegraphics[width=0.30\textwidth]{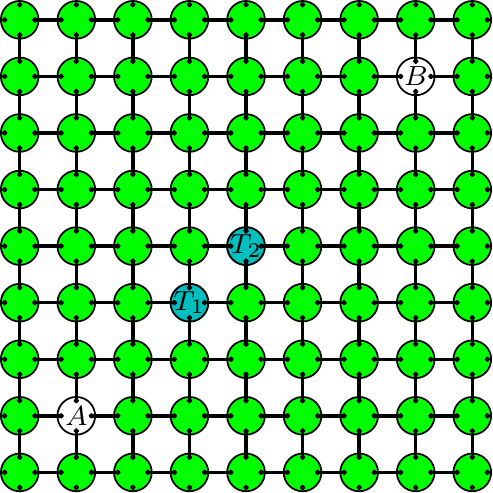}
		\label{fig:uneqdiag2}
	}
	\subfigure[Results for \textsc{diag-4-2-6} setting.]{%
		\includegraphics[width=0.30\textwidth]{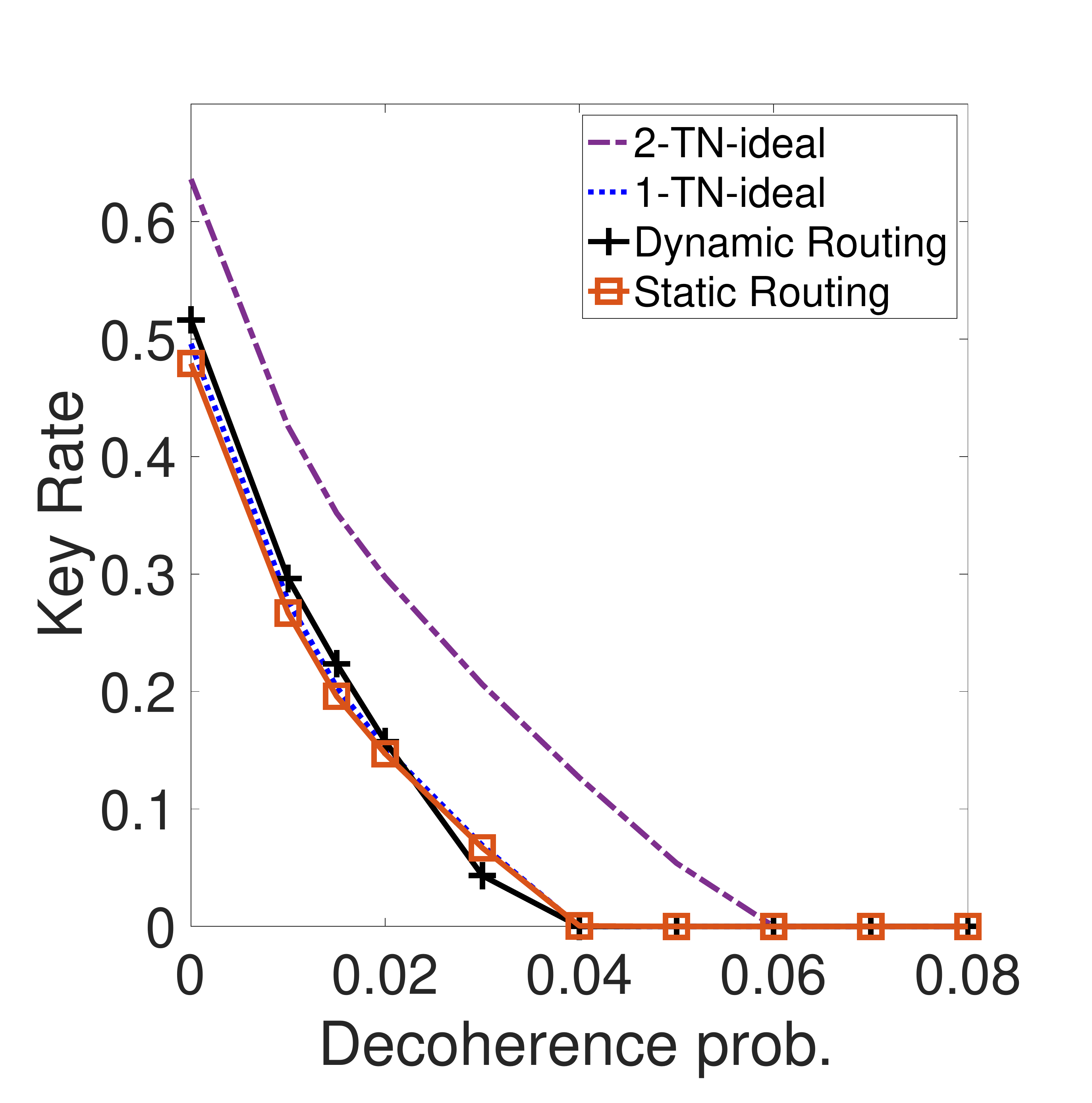}
		\label{fig:UneqDiagGlobal2}
	} \\
    \vspace{-0.1in}
	\subfigure[\textsc{Off-diag} setting.]{%
		\includegraphics[width=0.30\textwidth]{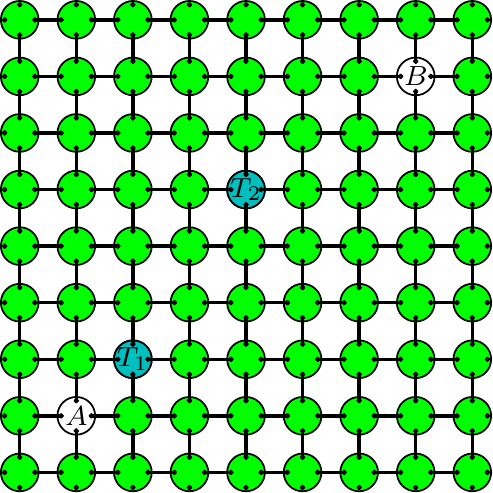}
		\label{fig:uneqdiag3}
	}
	\subfigure[Results for \textsc{Off-diag} setting.]{%
		\includegraphics[width=0.30\textwidth]{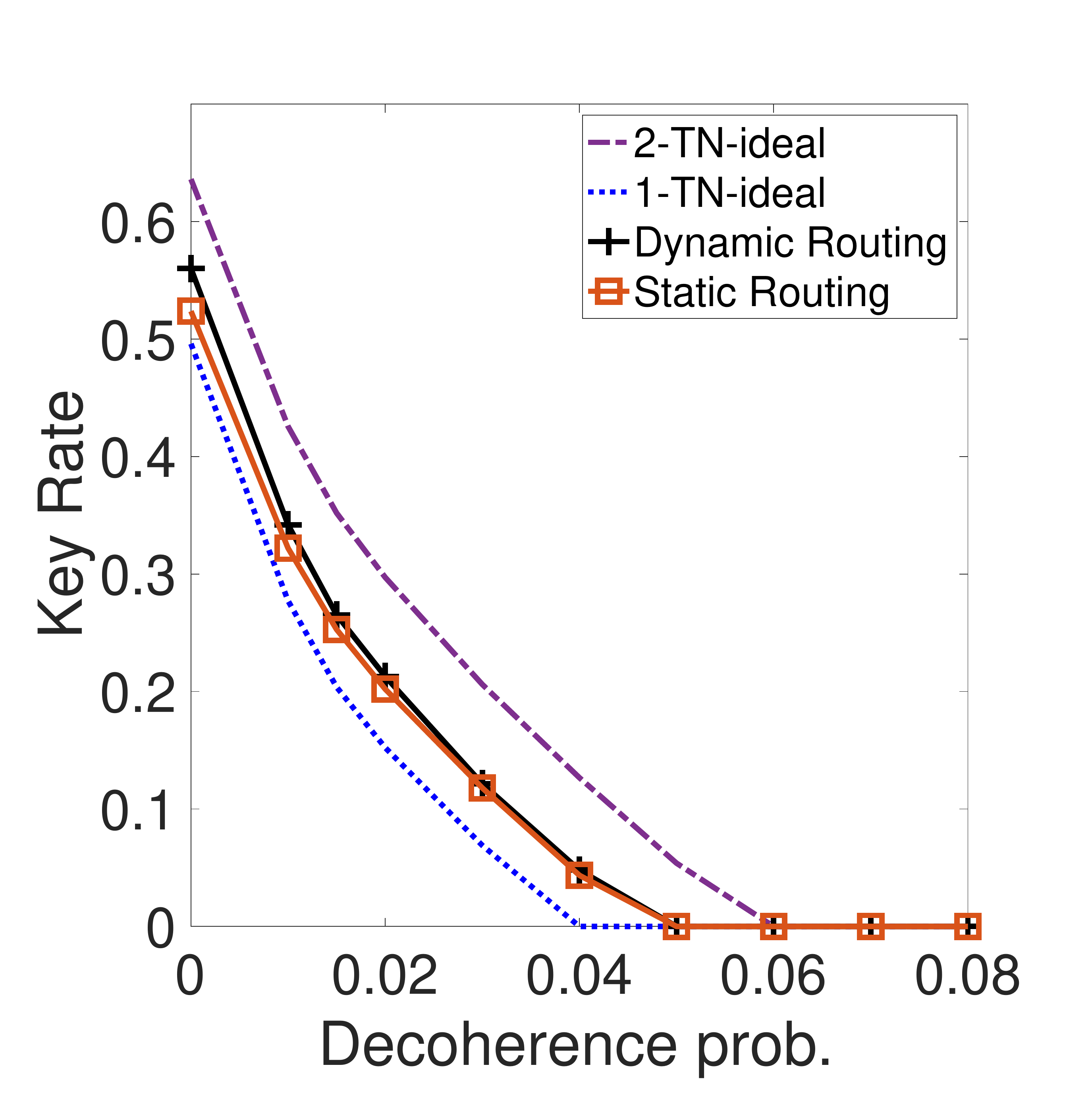}
		\label{fig:UneqDiagGlobal3}
	} 
    \vspace{-0.2in}
	\caption{Performance of dynamic routing in settings with two TNs with asymmetric placement. 
 }
    \label{fig:2-trusted}
\end{figure}
 



\smallskip
\noindent{\bf Two TNs.} We next consider several settings with two TNs under asymmetric placement, i.e., the distance between Alice and the TN that is closer to her  differs from the distance between Bob and the TN that is closer to him. Fig.~\ref{fig:2-trusted}a and c show two settings where the two TNs are placed along the diagonal; Fig.~\ref{fig:2-trusted}e shows one setting where one TN is on the diagonal, while the other is off the diagonal. We refer to the first setting as \textsc{diag-2-6-4}, where the three numbers represent the Manhattan distances between Alice and $T_1$, between $T_1$ and $T_2$, and between $T_2$ and Bob, respectively. Similarly, we refer to the second setting as  \textsc{diag-4-2-6}, and the ideal setting with two TNs as \textsc{diag-4-4-4}. In the third setting, $T_1$ is on the diagonal, with Manhattan distance 2 from Alice, and $T_2$ is off the diagonal, equidistant from both Bob and $T_1$ (both with Manhattan distance 5), and we refer to this setting as \textsc{off-diag}. 
Fig.~\ref{fig:2-trusted}b, d and f show the results of the above three settings. For reference, we also plot the key rate when using one TN or two TNs under ideal placement in these figures. 




The \textsc{diag-2-6-4} setting differs from the ideal setting with two TNs in that TN $T_1$ is closer to Alice, causing the distance between the two TNs, $T_1$ and $T_2$, to be larger. In this setting, we observe significant benefits of using dynamic routing over static routing. With no decoherence, the key rate of the dynamic routing is 0.531, more than twice of that under static routing (0.258), and also slightly higher than the ideal single TN scenario (which has key rate of 0.496). In general, dynamic routing achieves slightly higher key rate than the ideal single TN scenario for all the decoherence rates except when the the decoherence rate increases to 0.03, at which point dynamic routing leads to slightly lower key rate. This might be because at that decoherence rate, the longer paths that are selected due to dynamic routing become so noisy that the overall key rate is negatively affected. In Section~\ref{sec:dynamic-and-postprocess}, we will show that this effect can be mitigated by using CAD and our key-pool segmenting  techniques.

The \textsc{diag-4-2-6} setting differs from the ideal setting with two TNs in that one TN, $T_2$, is closer to the other TN, $T_1$, causing the distance  between $T_2$ and Bob to be larger. It is equivalent to the single TN ideal placement (i.e., $T_2$ is in the center of the grid), with an additional TN, $T_1$, very close to $T_2$. In this case, since the two TNs are so close to each other, they work similarly as a single node in the center. 
As a result, we observe much less benefit under dynamic routing over static routing (e.g., leading to 7.7\% improvement in key rate  when the decoherence rate is 0.0).
When the decoherence rate is below 0.03, the key rate of the single TN ideal placement is between those of dynamic routing and static routing. When the decoherence rate reaches 0.03, we again observe that dynamic routing leads to a slightly lower key rate than that of the single TN ideal placement; this effect will again be mitigated using key-pool segmenting and CAD (see later).

%

Last, we consider the \textsc{off-diag} setting. In this setting, we still observe that dynamic routing leads to higher key rate than static routing, particularly when {the} decoherence rate is low. As an example, when the decoherence rate is 0.0, the key rate under dynamic routing is 0.56, 7.7\% higher than that under under static routing (0.52), and only 17\% lower than the two TN ideal placement setting. Interestingly, in this setting, with the two TNs, both dynamic and static routing achieve significantly higher key rate than the ideal single trust node  scenario.   

Summarizing the above, when two TNs are placed  asymmetrically in the network, we see that the dynamic routing strategy can lead to significantly higher key rate than static routing in some cases. Since the placement of the two TNs is not ideal, the key rate even under  dynamic routing is lower than that when the two TNs are at ideal locations.
On the other hand, the key rate is significantly higher or similar to that under single TN ideal placement; the difference depends on the placement of the two TNs.

\subsection{Results of Classical Post-Processing}

We now evaluate the impact of classical post-processing, CAD and  
our key-pool segmenting technique, on key rate. 
Although 
CAD is a well understood tool, the 
interesting aspect here is its interactions with our key-pool segmenting technique in the network settings. For ease of exposition, we consider four settings that have no TN, a single TN at the ideal location, two TNs at their ideal locations, or the two corners of the grid, referred to as \textsc{no-TN}, \textsc{1TN-ideal}, \textsc{2TN-ideal}, and \textsc{2TN-corner}, respectively. For all these settings, it is sufficient to use static routing,  since they either have no TN or the TN(s) are placed at symmetric location(s). In Section~\ref{sec:dynamic-and-postprocess}, we consider dynamic routing together with these two  post-processing techniques for scenarios with asymmetric TN placement. 
Since both key-pool segmenting and CAD tend to bring benefits at high noise rates (and therefore at low key rates), we plot the graphs below on a log-linear scale to highlight the benefits of these two techniques. 


\begin{figure}[t]
	\centering
    \hspace{-.16in}
	\subfigure[\textsc{No-TN}.]{%
		\includegraphics[width=0.26\textwidth]{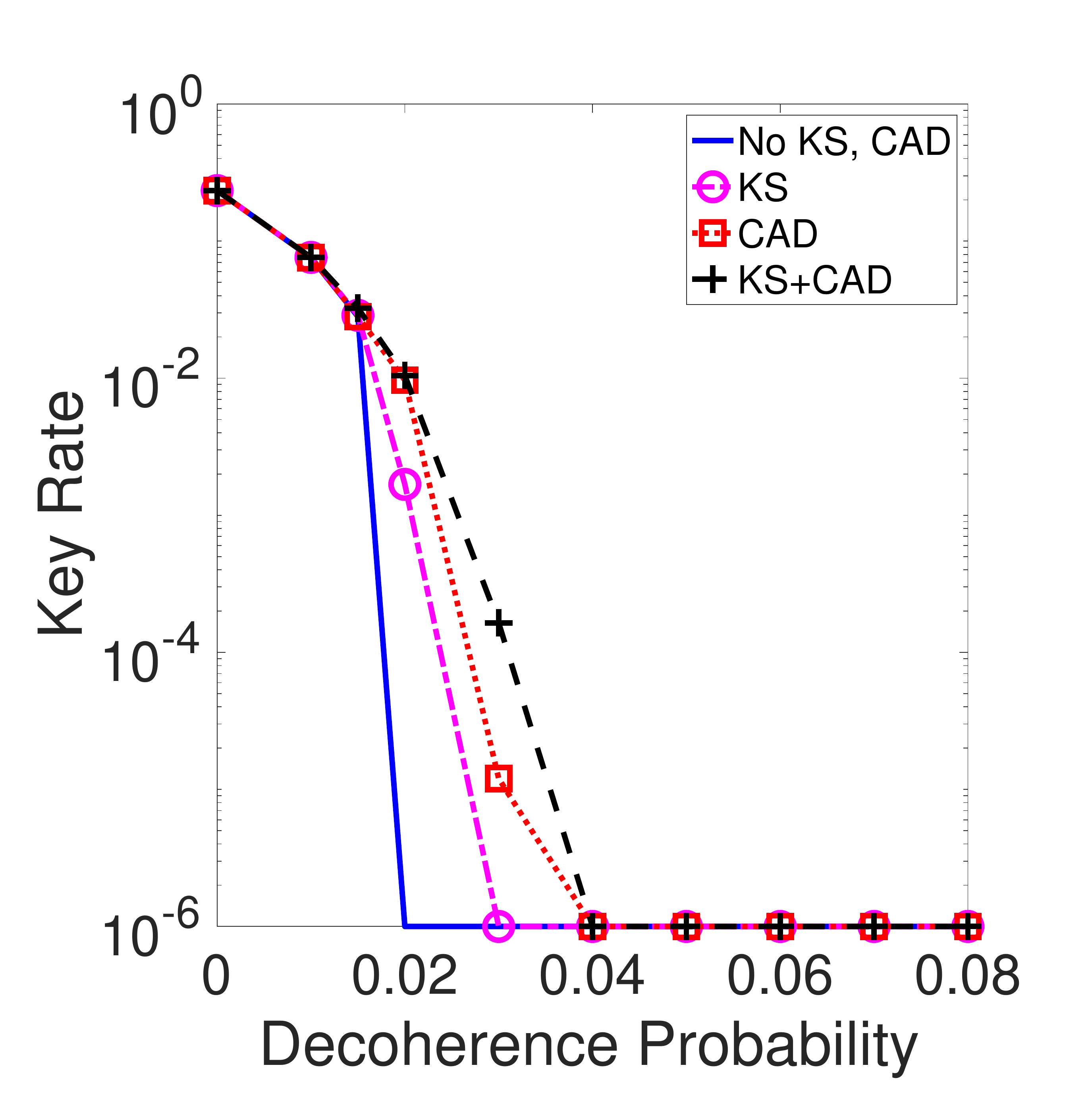}
	}
	\hspace{-.16in}
	\subfigure[\textsc{1TN-ideal}.]{%
		\includegraphics[width=0.25\textwidth]{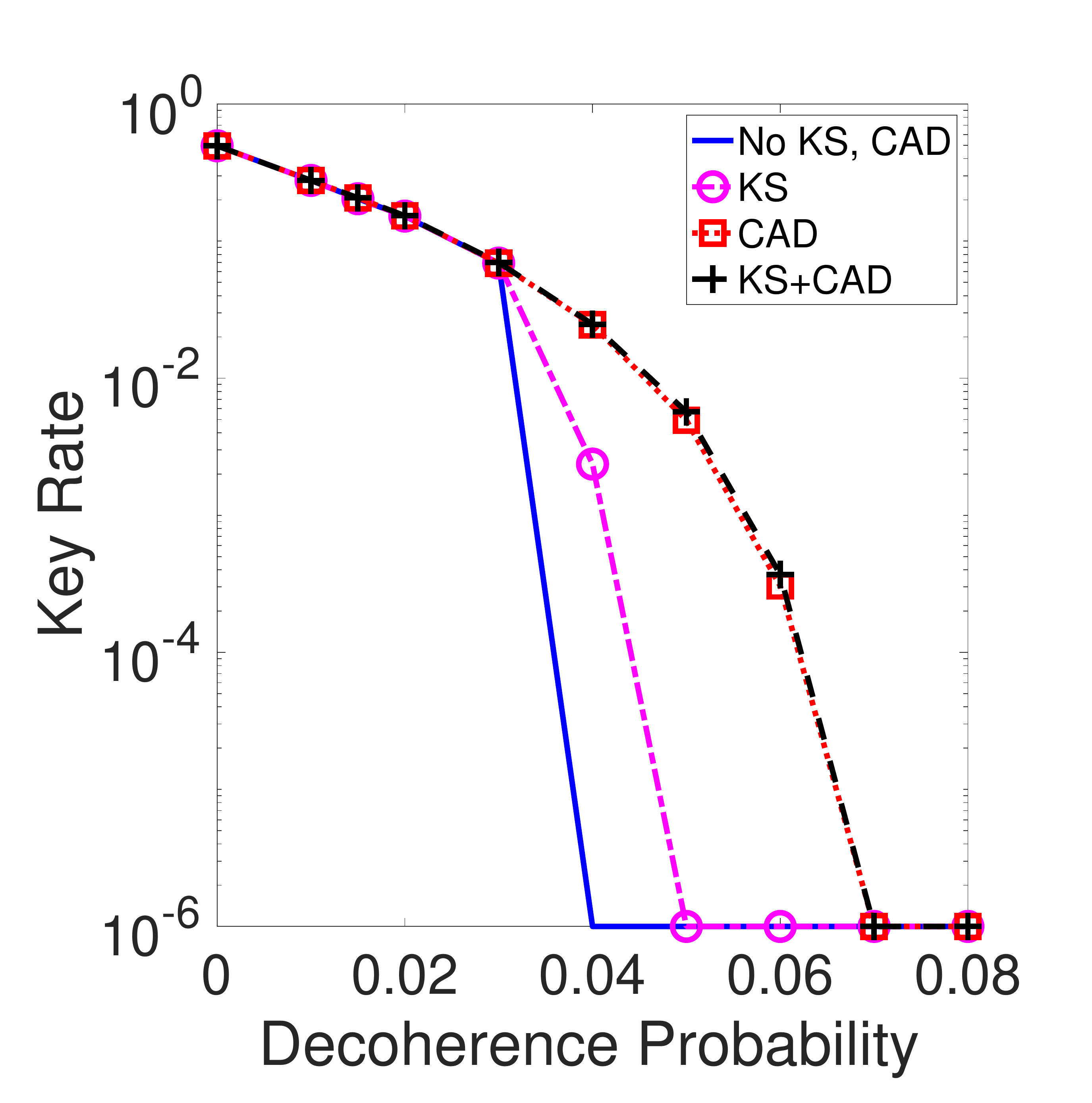}
	} 
    \hspace{-.16in}
	\subfigure[\textsc{2TN-ideal}.]{%
	\includegraphics[width=0.26\textwidth]{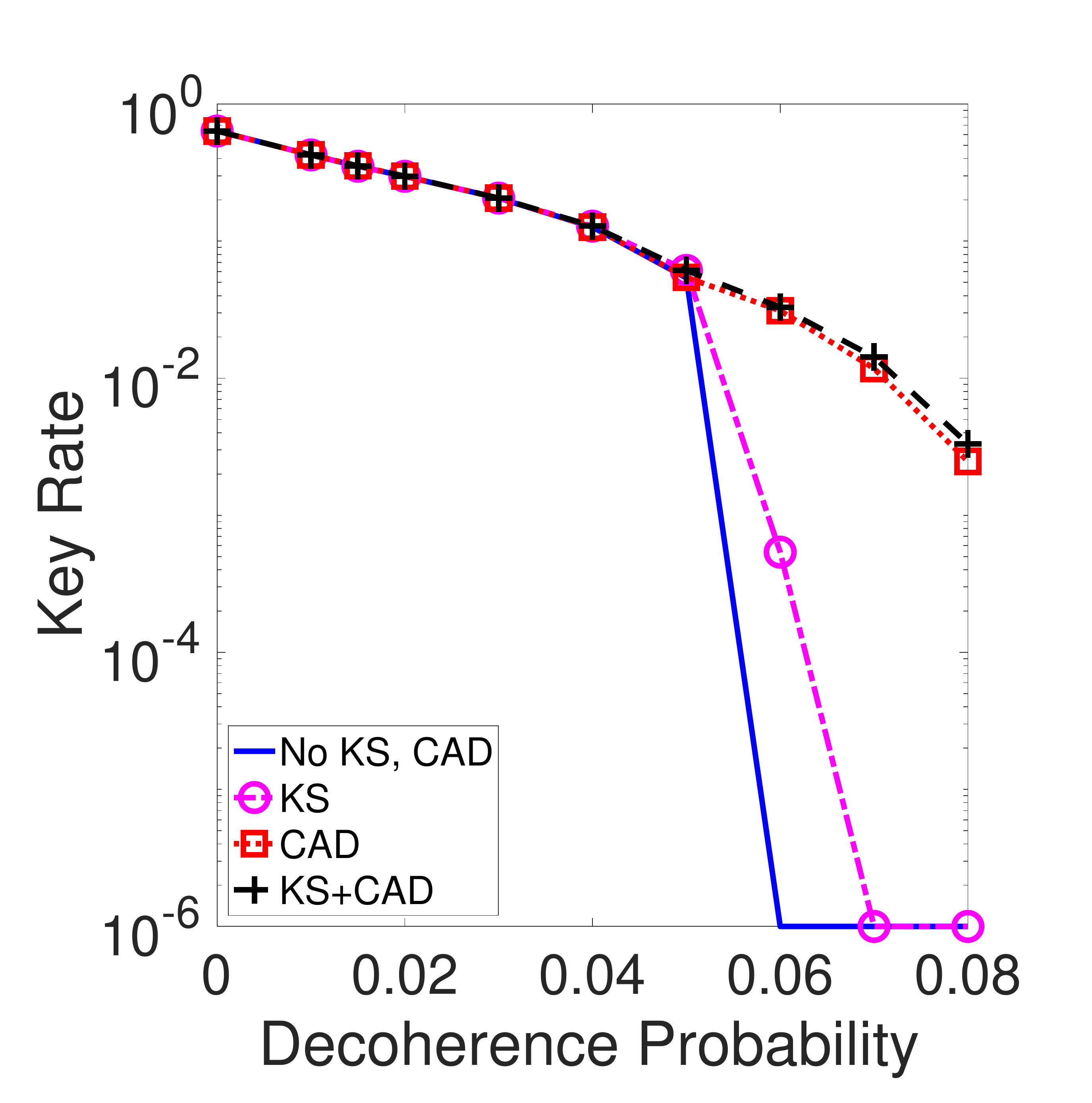}
	}
    \hspace{-.16in}
	\subfigure[\textsc{2TN-corner}.]{%
		\includegraphics[width=0.26\textwidth]{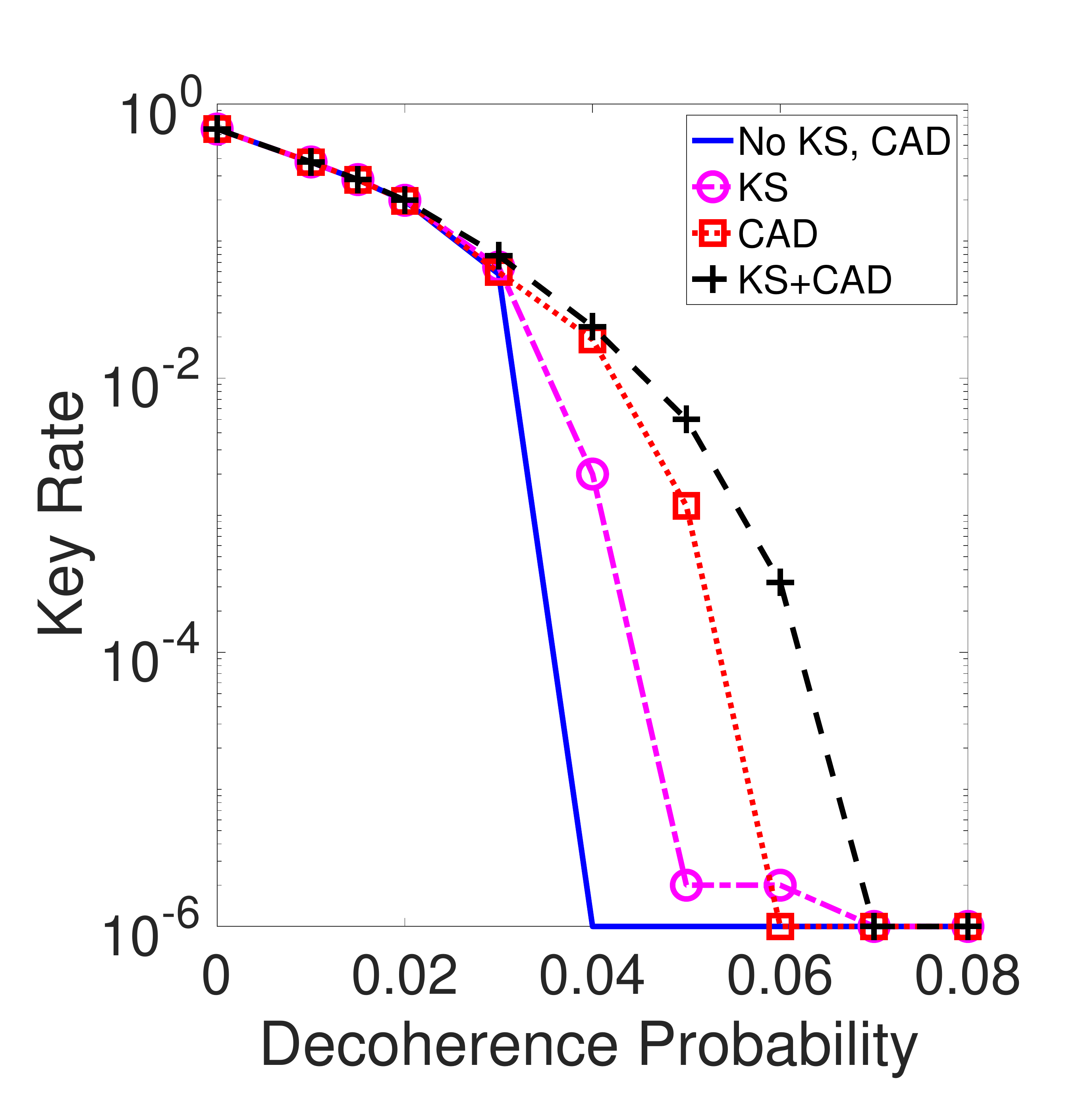}
	}
	\hspace{-.16in}
	\caption{Performance of CAD and key-pool segmenting techniques with static routing (which suffices for these settings). 
 }
	\label{fig:CADPooling}
\end{figure}

The \textsc{no-TN} setting  (Fig.~\ref{fig:CADPooling}a) serves as the baseline. In this setting, we see that both CAD and key-pool segmenting improve key rate when the decoherence rate is larger than 0.02. In addition, combining these two techniques can improve the key rate significantly than using them in isolation, e.g., leading to an order of magnitude improvement in key rate when the decoherence rate is 0.03. 

In \textsc{1TN-ideal} and \textsc{2TN-ideal} settings (Fig.~\ref{fig:CADPooling}b and c), while key-segmenting does improve key rate significantly when the decoherence rate is $\ge 0.03$, using key-segmenting and CAD together only brings slightly higher key rate 
beyond what is already provided by CAD. The observation is different in the \textsc{2TN-corner} setting (Fig.~\ref{fig:CADPooling}d), where we observe  significantly more benefits from combining key-pool segmenting and CAD over using each technique separately. The significantly higher benefits from 
key-pool segmenting beyond CAD in this setting might be because the TNs are at the two corners, and hence can lead to more diverse path lengths, resulting in more visible benefits from key-pool segmenting, which  differentiates paths of diverse qualities.

\subsection{Results of Dynamic Routing with Classical Post-Processing} \label{sec:dynamic-and-postprocess}

\begin{figure}[t]
	\centering
    \hspace{-.16in}
	\subfigure[\textsc{off-center}.]{%
		\includegraphics[width=0.26\textwidth]{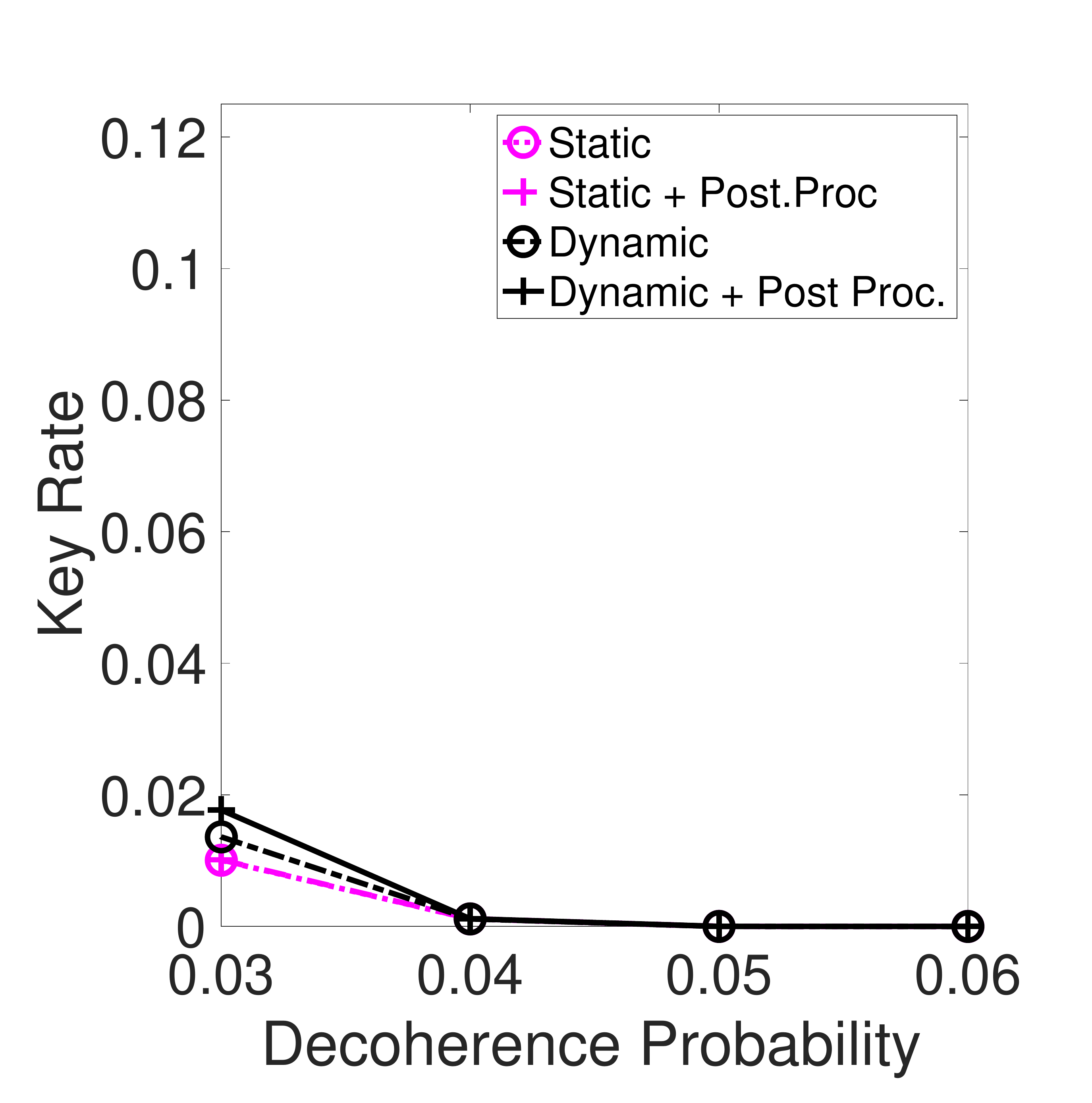}
	}
	\hspace{-.16in}
	\subfigure[\textsc{diag-2-6-4}.]{%
		\includegraphics[width=0.26\textwidth]{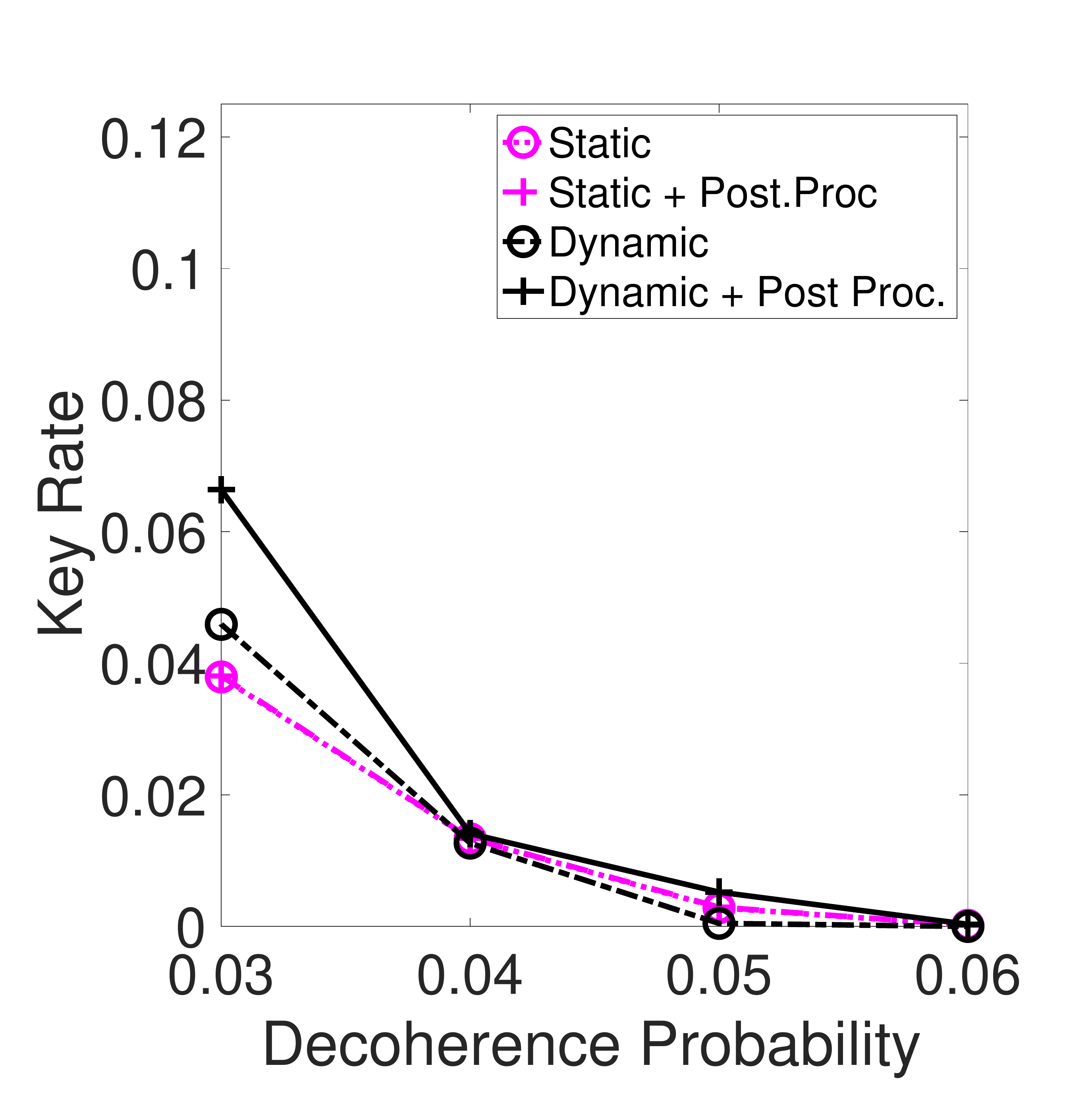}
	}
    \hspace{-.16in}
    \subfigure[\textsc{diag-4-2-6}.]{%
		\includegraphics[width=0.26\textwidth]{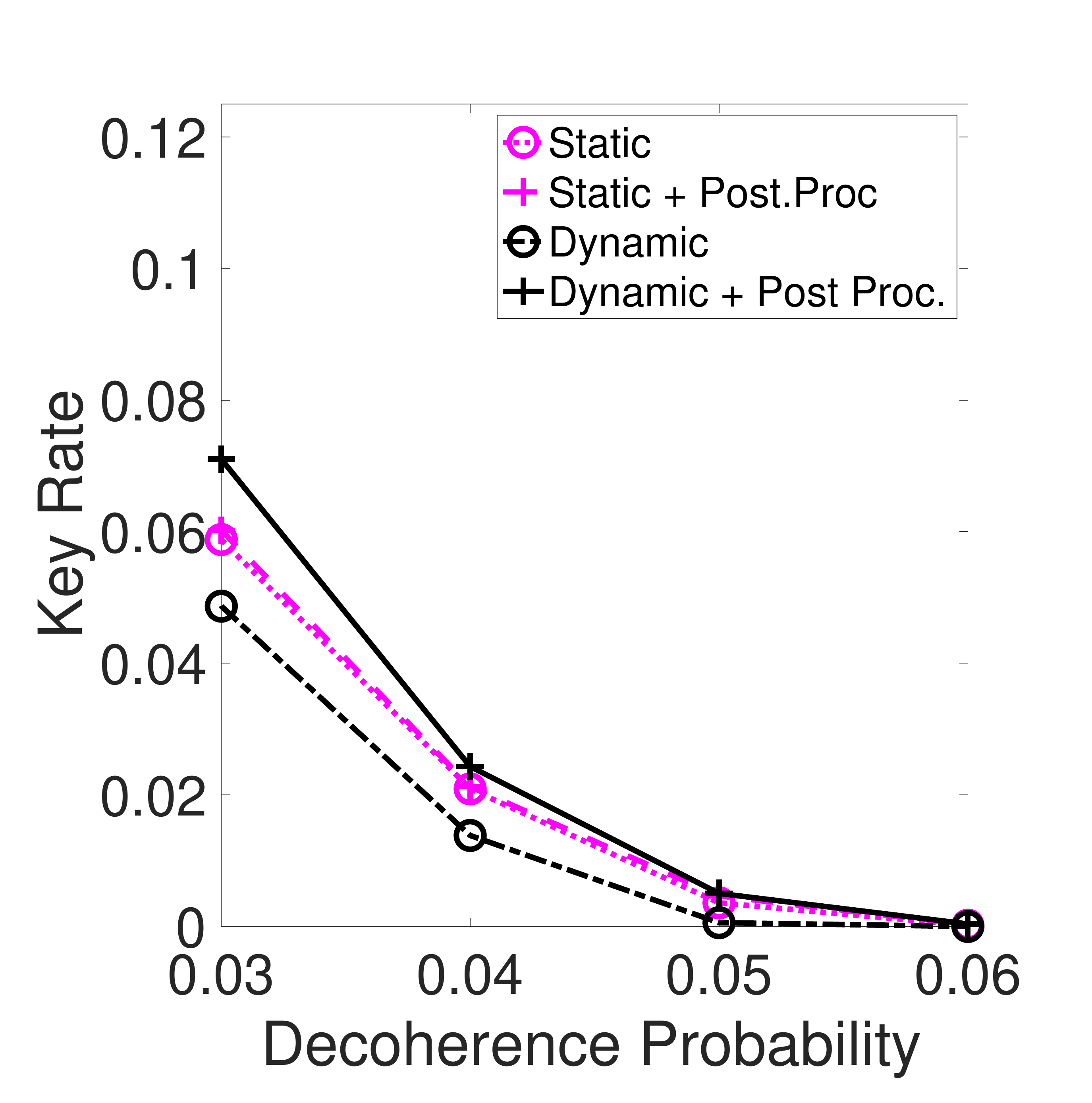}
	} 
	\hspace{-.16in}
	\subfigure[\textsc{off-diag}.]{%
		\includegraphics[width=0.26\textwidth]{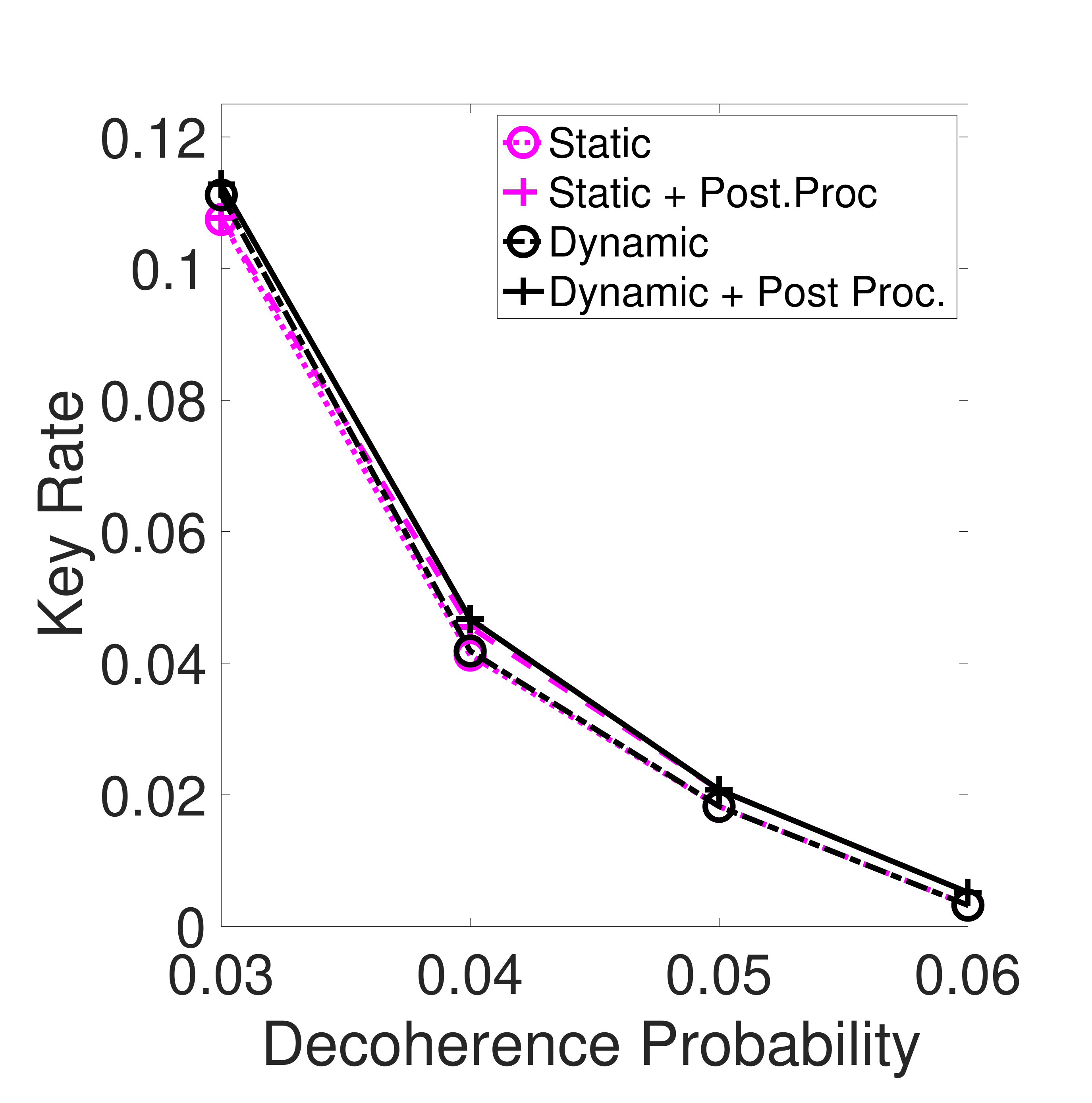}
	}
    \vspace{-0.2in}
	\caption{Key rate when combining dynamic routing and post-processing techniques. 
 }\label{fig:diffBoth}
\end{figure}

We now evaluate the benefits of combining dynamic routing and post-processing methods (both key-pool segmenting and CAD) for the settings with asymmetric TN placement. Specifically, we consider four asymmetric settings that we considered earlier: one with a single TN, i.e.,  \textsc{off-center}, the other three with two TNs, i.e.,  \textsc{diag-2-6-4}, \textsc{diag-4-2-6}, and \textsc{off-diag}. Since post-processing techniques only bring benefits under high noise rates, Fig.~\ref{fig:diffBoth} only plots the key rate when the decoherence rate is $\ge 0.03$. We see that for all the settings, adding post-processing to the raw key pools obtained from dynamic routing improves the key rate 
(comparing the solid and dashed black lines), and the benefits over using dynamic routing only 
 are particularly clear under \textsc{diag-2-6-4} and \textsc{diag-4-2-6}. The above benefits are expected since dynamic routing can lead to much longer paths, and higher noise rate, which can benefit from post-processing techniques. 
For comparison, Fig.~\ref{fig:diffBoth} also plots the key rate under static routing,  with and without post-processing. As expected, the benefits of using post-processing techniques become much less noticeable with static routing, since shorter paths are chosen with higher priority at all times under static routing.

\section{Conclusion and Future Work}
In this paper, we consider  
QKD networks consisting of quantum repeaters and a small number of TNs. Such networks serve as a middle ground between current QKD networks used today (which consist of only TNs) and the future true Quantum Internet. We have developed dynamic routing strategies that make routing decisions based on the current network state; we also evaluated the effects of classical/quantum post-processing strategies in this network setting. Using simulations, we show that the dynamic routing strategy can improve the key rate between two users significantly in certain settings, particularly with asymmetric TN placement (which is likely to occur in real-world deployments of such networks).  The post-processing techniques can significantly improve key rate in high noise scenarios. Furthermore, combining the dynamic routing strategy and the post-processing strategies can further improve the overall performance of the QKD network.  

In this work, we have combined dynamic routing  and post-processing techniques with global-state routing (i.e., the link-level entanglement state is known to Alice, Bob and all TNs). As future work, we will explore using these two types of techniques with local routing protocols, where only the entanglement state between neighboring nodes are known, which is more realistic in large-scale networks. Other future directions include investigating networks of different topologies and settings with multiple end users,  as well as developing efficient optimal routing strategies for such near-future networks. Finally, an important future problem is studying the effect of the network in the finite key setting, as opposed to looking only at asymptotic trends as we did here.  In the finite key setting, one needs to take into account various effects such as imperfect sampling.  How our strategies perform in such scenarios remains an interesting open question.

\section*{Disclaimer}
This paper was prepared for information purposes by the teams of researchers from the various institutions
identified above, including the Global Technology Applied Research group of JPMorgan
Chase Bank, N.A.. This paper is not a product of the Research Department of JPMorgan Chase \& Co. or its
affiliates. Neither JPMorgan Chase \& Co. nor any of its affiliates make any explicit or implied representation
or warranty and none of them accept any liability in connection with this paper, including, but limited to,
the completeness, accuracy, reliability of information contained herein and the potential legal, compliance, tax
or accounting effects thereof. This document is not intended as investment research or investment advice, or
a recommendation, offer or solicitation for the purchase or sale of any security, financial instrument, financial
product or service, or to be used in any way for evaluating the merits of participating in any transaction.

\balance

\end{document}